\documentclass[]{elsarticle}

\usepackage[margin=2.3cm]{geometry} %minimize the white space at the edges
\usepackage{amstext,amsmath,amssymb}
\usepackage{hyperref} %is always last to be included

\journal{Chemical Engineering Science}
\newcommand{\EE}{\ensuremath{\mathcal{E}}}

\begin{document}

%%%%%%%%%%%%%%%%%%%%%%%
\begin{frontmatter}
%%%%%%%%%%%%%%%%%%%%%%%

\title{Small-scale resolving simulations of the turbulent mixing in confined planar jets using one-dimensional turbulence} 

%%%%%%%%%%%%%%%%%%%%%%%

%%%% Group authors per affiliation:
%\author{Marten~Klein\corref{cor1}}
%\ead{marten.klein@b-tu.de}

%\author{Christian~Zenker}
%\ead{christian.zenker@b-tu.de}

%\author{Heiko~Schmidt}
%\ead{heiko.schmidt@b-tu.de}

%\address{Lehrstuhl Numerische Str\"omungs- und Gasdynamik, Brandenburgische Technische Universit\"at Cottbus-Senftenberg, Siemens-Halske-Ring 14, D-03046 Cottbus, Germany}

%%%% Or include affiliations in footnotes:
\author[addr1]{Marten Klein\corref{cor1}}
\ead{marten.klein@b-tu.de}
\cortext[cor1]{Corresponding author.
  Tel.: +49-(0)355-695-127;
  Fax: +49-(0)355-694-891.
}

\author[addr1]{Christian Zenker}
%\ead{christian.zenker@b-tu.de}

\author[addr1]{Heiko Schmidt}
%\ead{heiko.schmidt@b-tu.de}

\address[addr1]{Lehrstuhl Numerische Str\"omungs- und Gasdynamik, Brandenburgische Technische Universit\"at (BTU) Cottbus-Senftenberg, Siemens-Halske-Ring 14, D-03046 Cottbus, Germany}

%%%%%%%%%%%%%%%%%%%%%%%

\begin{abstract}
Small-scale effects of turbulent mixing are numerically investigated by applying the map-based, stochastic, one-dimensional turbulence (ODT) model to confined planar jets.
The model validation is carried out for the momentum transport by comparing ODT results to available reference data for the bulk Reynolds numbers $Re=20\,000$ and $40\,000$.
Various pointwise statistical quantities are computed and compared to the available reference data.
We show that these quantities can be captured well, or at least to a reasonable extent, by the stand-alone model formulation and for fixed model parameters.
Only the root-mean-square velocity fluctuations remain systematically underestimated in the ODT results (by an approximate factor of $1.5$).
Afterwards, the turbulent transport of a passive scalar is addressed for the Schmidt numbers $Sc=1$ and $1250$.
For the high Schmidt number and in contrast to the velocity fluctuations, it is shown that the scalar fluctuation variance is up to ten times larger in the ODT simulations resolving the Batchelor scale.
The fluctuation variance is notably smaller for the lower Schmidt number, but exhibits better agreement with the references at a nominally higher Schmidt number.
We suggest that this is due to implicit filtering in the references, which barely resolve the Kolmogorov scale.
ODT turbulence spectra support this interpretation since a Batchelor-like scalar turbulence spectrum is only observed for the higher Schmidt number.
With the aid of these spectra and the fluctuation statistics we conclude that implicit filtering has a similar effect as a reduction of the Schmidt number.
\end{abstract}


%%%%%%%%%%%%%%%%%%%%%%%
\end{frontmatter}
%%%%%%%%%%%%%%%%%%%%%%%

%\linenumbers

%\tableofcontents %only for draft

%%%%%%%%%%%%%%%%%%%%%%%
\section{Introduction} \label{sec:intro}
%%%%%%%%%%%%%%%%%%%%%%%

Turbulent mixing denotes the redistribution of mass, momentum, energy or chemical species due to an unsteady, three-dimensional flow characterized by a range of time and length scales.
It it is well known that, from a qualitative point of view, the presence of turbulence significantly enhances the transport across a fluid layer compared to pure molecular diffusion.
For a given application, one also wishes to gain a quantitative understanding of the effects.
This, however, cannot always be achieved, or dealt with, in a straightforward and simple way.

A major obstacle in this sense is the extent of the turbulent scales that need to be resolved.
This extent is expressed by the (bulk) Reynolds number $Re=UL/\nu$, where $U$ is the bulk velocity, $L$ the integral length scale, and $\nu$ the kinematic viscosity of the working fluid.
The viscous cut-off occurs at the Kolmogorov length scale $\eta_{K}\sim Re^{-3/4} L$~\cite{Kolmogorov:1941}.
A fully-resolved, three-dimensional direct numerical simulation (3-D DNS) demands $O(Re^{9/4})$ grid cells.
The maximum Reynolds number is therefore at present limited to $Re\lesssim10^5$ for idealized configurations such as channel flows (e.g.~\cite{Lee_Moser:2015, Brethouwer:2018}).
The resolution requirements can become even more restrictive when scalars have to be taken into account, for example, the temperature, chemical species, tracers, dye or aerosols.
These demand the resolution of the Batchelor scale $\eta_{B}=Sc^{-1/2} \eta_{K}$~\citep{Batchelor:1959}, where $Sc=\nu/\Gamma$ is the Schmidt number and $\Gamma$ the scalar diffusivity.
For flows with large characteristic Schmidt and Reynolds numbers, the computational requirements for resolution of all scales soon becomes too restrictive, therefore demanding the use of turbulence modeling.

Common modeling approaches are based on Reynolds-averaged Navier--Stokes~(RANS) simulations or large-eddy simulations~(LES).
For the former, the general idea dates back to Boussinesq (eddy viscosity hypothesis)~\cite{Boussinesq:1903} and for the latter to Smagorinsky~\cite{Smagorinsky:1963}, who suggested the modeling of the unresolved turbulent momentum fluxes $R_{ij}=\langle u_i'u_j'\rangle$ by diffusive fluxes $-2 \nu_t S_{ij}$ of the resolved scales, where $S_{ij}$ is the filtered strain-rate tensor and $\nu_t$ the turbulent eddy-viscosity. 
This viscosity is not a fluid property and needs to be determined with the aid of additional closure equations and additional assumptions (see e.g.~\cite{Ferziger_Peric:2002} for an overview).
An alternative modeling approach is given by the map-based, stochastic, one-dimensional turbulence (ODT) model.
In ODT, numerical efficiency is obtained by a lower-order formulation.
Accuracy, however, is addressed by resolving all scales within a quasi-one-dimensional framework.
In comparison to RANS or LES modeling, there is no need for closure and, thus, no eddy viscosity or turbulent Schmidt number involved.
The ODT model has been validated from a fundamental point of view with applications in a variety of canonical flow problems.
Among others, there have been applications to isotropic turbulence (e.g.~\cite{Kerstein:1999, Giddey_etal:2018}), free shear flows (e.g.~\cite{Kerstein_etal:2001, Ashurst_Kerstein:2005}), turbulent convection (e.g.~\cite{Dreeben_Kerstein:2000, Wunsch_Kerstein:2005}), boundary layers (e.g.~\cite{Kerstein_Wunsch:2006, Fragner_Schmidt:2017}), and chemically reacting flows (e.g.~\cite{Echekki_etal:2001, Hewson_Kerstein:2001, Monson_etal:2016}).
Latter are specifically challenging as they demand an accurate representation of the small-scale turbulent mixing.
This might explain why ODT has been used increasingly over the last years to study chemistry-turbulence interactions.
It is in this context rather surprising that the transport of passive scalars has not been considered in as much detail~\cite{Ashurst_etal:2003}, despite its fundamental relevance for this kind of application.
This circumstance has, in fact, been recognized only recently (e.g.~\cite{Giddey_etal:2018, Klein_Schmidt:2017, Klein_Schmidt_PAMM:2018}).

\begin{figure}[t]
  \centering
  \includegraphics[scale=0.4]{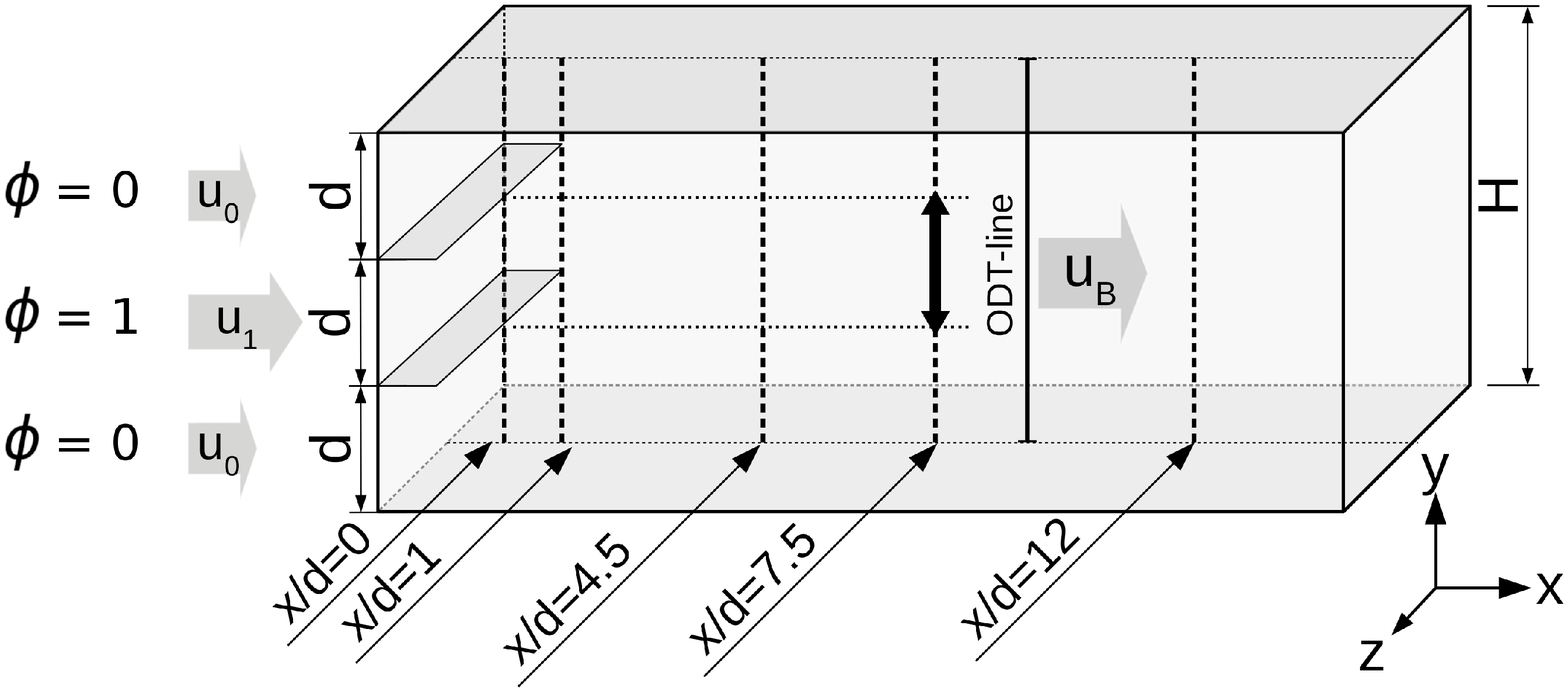}
  \caption{%
    Schematic of the confined planar jet (not to scale).
    The coordinates $(x,y,z)$ denote the streamwise, wall-normal and spanwise directions.
    Three small channels with height $d$, free-stream velocities $U_0$; $U_1$; $U_0$, and a passive scalar with mass fractions $\phi=0$; $1$; $0$ enter a large channel of height $H=3d$.
    ODT simulations are conducted the central $(x,y)$-plane of the large channel with a moving ODT line, which is advected with the bulk velocity $U_B=(2U_0+U_1)/3$.
    Statistical quantities are compared with available reference data at the marked locations~$x/d$.
    Turbulence spectra are computed for the central thick black interval.
  }
  \label{img:scheme_reactor}
\end{figure}

In this work, we apply ODT to numerically investigate confined planar jets.
The flow configuration is sketched in figure~\ref{img:scheme_reactor} and forms a well-defined, canonical flow problem, which is representative of a variety of flows in chemical engineering applications.
Here, we mainly aim to address the effect of insufficient resolution due to either numerical resources or measurement equipment limitations.
The dynamically adaptive ODT implementation~\cite{Lignell_etal:2013} used allows to resolve the flow down to the Kolmogorov and Batchelor scales for $Sc=1$ and $1250$.
The scalar with $Sc=1$ is considered here as a comparison to mimic the truncation of the resolution at around the Kolmogorov scale, similar to the available reference data~\cite{Feng2005, Kong2012, Arshad_etal:2018}.
This is possible here due to the mainly dissipative nature of the small-scales.
By comparing ODT results to these reference data, we show that the fluctuations of a high-$Sc$ scalar can increase drastically (up to a factor $10$) once the grid resolves the Batchelor scale.
The mean scalar distribution, however, does not differ much even from coarse-resolution RANS results.
This suggests that, in the hierarchy of models, ODT is an interesting candidate for applications in which small-scale resolution is critical but DNS are not feasible.

The rest of this paper is organized as follows:
In section~\ref{sec:model}, a description of the ODT model is given.
In section~\ref{sec:config}, the application of ODT to the confined jet is discussed.
In section~\ref{sec:ic}, the generation of turbulent inflow conditions is described.
In section~\ref{sec:results}, ODT results are compared to available reference data in terms of various low-order statistical quantities.
This is done first for the velocity and then for the passive scalar.
At the end, we also discuss turbulence spectra for the velocity and the passive scalar.
In section~\ref{sec:conc}, we conclude our main results.
Additional material relevant to reproduce the results is given in the Appendices. 
\ref{sec:stats} details the computation of point-wise statistical quantities in ODT.
\ref{sec:comp-spectra} comments on the procedure of computing one-dimensional turbulence spectra.
Finally, \ref{sec:ic-odt} addresses the consistency of the turbulent inflow conditions.

%%%%%%%%%%%%%%%%%%%%%%%
\section{ODT model formulation} \label{sec:model}
%%%%%%%%%%%%%%%%%%%%%%%

The ODT model aims to represent the evolution of the velocity vector and scalar fields along a line.
One-dimensional, deterministic, diffusion equations are discretized and solved numerically along this line while the effect of Navier--Stokes turbulence is modeled by stochastic eddy events sampled in a discrete-in-time fashion.
In the present work, the ODT domain corresponds to a wall-normal line that is displaced uniformly with the bulk velocity.
The aim is to capture the main features of the spatio-temporal evolution of the flow using ODT as stand-alone tool. 
In the following, we give a short but complete description of the ODT model based on the formulation introduced by Kerstein et al.~\cite{Kerstein:1999,Kerstein_etal:2001}.

%%%%%%%%%%%%%%%%%%%%%%
\subsection{Governing equations} \label{sec:eqs}
%%%%%%%%%%%%%%%%%%%%%%%

We consider isothermal constant-density flows with a single passive scalar without sources or sinks.
The scalar is prescribed by the initial condition and then redistributed by turbulent and molecular diffusion.
The continuity equation for the velocity and the Poisson equation for the pressure are not included in the model due to its one-dimensional nature.
Nevertheless, the effects of the fluctuating pressure-gradient forces are taken into account in the stochastic part of the model as will be discussed below. 
Altogether, the ODT governing equations~\cite{Kerstein_etal:2001} are given by
\begin{align}
  \frac{\partial u_i}{\partial t} + \EE_i(u_j) &= \dfrac{\partial}{\partial y} \left( \nu \frac{\partial u_i}{\partial y} \right) + \dfrac{f_i}{\rho},
  \label{eq:gov1}
  \\
  \frac{\partial \phi}{\partial t} + \EE_\phi(u_j) &= \dfrac{\partial}{\partial y} \left( \Gamma \frac{\partial \phi}{\partial y} \right), 
  \label{eq:gov2}
\end{align}
where $u_i$ denotes the velocity vector $(u_i)=(u,v,w)$, $\phi$ the concentration of the passive scalar, $f_i$ the external forces driving the flow, $\rho$ the constant fluid density, and $\EE$ the stochastic eddy events.
Other quantities have been introduced earlier.

The stochastic terms $\EE$ in equations~\eqref{eq:gov1} and \eqref{eq:gov2} are zero in between any two instantaneous eddy events.
Between any two such perturbations, a set of one-dimensional diffusion equations is solved continuously in time.
The spatial discretization along the ODT line is done with a finite-volume method on an adaptive grid~\cite{Lignell_etal:2013}.
The velocity vector and the passive scalar are located at the cell centers of the same grid. 
This is permissible in 1-D and aids the conservation properties of the scheme.
The minimum allowed cell size $\Delta y_{min}$, which is needed for dynamic re-meshing, is of the order of the Batchelor scale $\eta_B\simeq Sc^{-1/2} Re^{-3/4} H$.
Similarly, the maximum allowed cell size $\Delta y_{max}$ is $H/50$ for case~A and $H/1000$ for case~B in order to minimize the numerical transport and implicit filtering of scalar fluctuations.
The temporal discretization is done here with an explicit Euler method.
This is not a limitation but since any higher-order accuracy of the deterministic part would still result in a first-order global scheme due to the instantaneous implementation of the stochastic mapping events.

%%%%%%%%%%%%%%%%%%%%%%%
\subsection{Formulation of the eddy events} \label{sec:eddies}
%%%%%%%%%%%%%%%%%%%%%%%

Eddy events enter in equations~\eqref{eq:gov1} and \eqref{eq:gov2} as a stochastic term that manifests itself by instantaneous modifications of the flow variables.
Two mathematical operations are used to represent the effects of turbulent advection and pressure fluctuations.
When an eddy event is selected, the variables at location $y$ are instantaneously replaced by the values at mapped location $f(y)$.
For the scalar and the velocity vector, these operations~\cite{Kerstein_etal:2001} are given by
\begin{align}
  \mathcal{E}_{\phi}&: & \phi(y) &\to \phi''(y) = \phi\big(f(y)\big),
  \label{eq:eddy1}
  \\
  \mathcal{E}_{i}&: & u_i(y) &\to u_i''(y) = u_i\big(f(y)\big) + c_i\, K(y),
  \label{eq:eddy2}
\end{align}
where $f(y)$ denotes a mapping function, $K(y)=y-f(y)$ a kernel function, and $c_i$ the coefficients representing the effect of pressure fluctuations.
These coefficients control the redistribution of the kinetic energy among the velocity components.

In Navier--Stokes turbulence, the turnover of a single eddy increases locally the gradients of the flow variables on the length scale of that eddy.
This notion is addressed in ODT by using the triplet map as a mapping function~\cite{Kerstein:1999}.
For an eddy event of size $l$ occurring at location $y_0$, the mapping takes place for the interval $[y_0,y_0+l]$ and is given by
\begin{equation}
  f(y) = y_0 + \left\{
  \begin{array}{ll}
         3(y-y_0) &\ \mathrm{for}\ y-y_0 \in \big[    0,  l/3 \big], \\%
    2l - 3(y-y_0) &\ \mathrm{for}\ y-y_0 \in \big[  l/3, 2l/3 \big], \\%
    3(y-y_0) - 2l &\ \mathrm{for}\ y-y_0 \in \big[ 2l/3,  l   \big], \\%
      y-y_0       &\ \mathrm{otherwise.}
  \end{array}
  \right.
  \label{eq:triplet}
\end{equation}

In the dynamically adaptive formulation, profiles of the flow variables are compressed to one third of their length, copied twice to fill the eddy size interval and with the central copy flipped in order to ensure continuity~\cite{Lignell_etal:2013}.
Irrespective of the implementation details, the important properties of the triplet map are that it is (i)~measure-preserving and (ii)~does not introduce discontinuities along the ODT line.
These two aspects are important for the conservation properties of the method.
Note that, due to the triplet map, kinetic energy is brought from large to small scales in a scale-local fashion, which results in a direct energy cascade as in 3-D turbulence~\cite{Kerstein:1999}.

The last term in equation~\eqref{eq:eddy2} models the effect of fluctuating pressure gradient forces.
The kernel function $K(y)$ is a measure for the map-induced fluid displacement and the coefficient vector $c_i$ for the efficiency of the inter-component kinetic energy transfer.
The change of the kinetic energy in the $i$th velocity component due to the application of an eddy event is given by
\begin{equation}
 \Delta E_i = \dfrac{1}{2l} \int_{y_0}^{y_0+l} \left[ \big( u_i(f(y)) + c_i K(y) \big)^2 - u_i^2(y) \right] \,\mathrm{d}y.
 \label{eq:dE}
\end{equation}

Energy conservation requires that the sum of the changes vanishes, $\Delta E_1 + \Delta E_2 + \Delta E_3 = 0$~\cite{Kerstein_etal:2001}.
The $c_i$ are obtained by a maximization of the inter-component kinetic energy transfer (i.e.~$-\Delta E_i$) with respect to $c_i$.
This yields
\begin{equation}
   c_i = \frac{1}{K_K}
     \left[ - u_{K,i} + \mathrm{sgn}(u_{K,i}) \sqrt{ (1-\alpha) u_{K,i}^2 + \frac{\alpha}{2} \left( u_{K,j}^2 + u_{K,k}^2\right) } \,\right],
  \label{eq:ci}
\end{equation}
where $u_{K,i}=\int u_i\big(f(y)\big)\, K(y) \,\mathrm{d}y$ denotes the kernel-weighted velocity vector, $K_K=\int K^2(y) \,\mathrm{d}y$ the squared kernel, which is related to the map-induced fluid displacement, the indexes $(ijk)$ are permutations of $(123)$, and $\alpha$ is a model parameter that controls the efficiency of the inter-component energy transfer due to fluctuating pressure gradient forces. 
The parameter $\alpha$ specifies the fraction of the available (extractable) kinetic energy that is actually used for redistribution~\cite{Kerstein_etal:2001}.
It takes values between $0$ (no redistribution) and $1$ (maximal redistribution).
Here we select $\alpha=2/3$ assuming a tendency to small-scale isotropy, that is, a relaxation to locally homogeneous isotropic turbulence~\cite{Kerstein_etal:2001}.

%%%%%%%%%%%%%%%%%%%%%%%
\subsection{Stochastic eddy selection} \label{sec:selection}
%%%%%%%%%%%%%%%%%%%%%%%

Eddy events are characterized by three random variables: the eddy size $l$, the position $y_0$, and their time $t$ of occurrence.
In theory, these variables can be sampled from the eddy-rate distribution $\lambda$, whereby $\lambda(l, y_0, t)\,\mathrm{d}l\,\mathrm{d}y_0\,\mathrm{d}t$ gives the number of eddy events in the size range $[l, l+\mathrm{d}l]$, the position range $[y_0, y_0 + \mathrm{d}y_0]$ and the time interval $[t,t+\mathrm{d}t]$.
This distribution, however, depends on the flow state itself and is therefore unknown.

In practice, the repeated and costly construction of $\lambda$ is replaced by a more efficient thinning-and-rejection method~\cite{Kerstein:1999}.
For this purpose, the eddy-rate distribution is re-written using dimensional arguments and the fact that $y_0$ is related to the turbulence region of the flow but not its turbulence properties.
With $l$ as length scale, we arrive at $\lambda=C\,\tau^{-1}\,l^{-2}$, where $C$ is a proportionality constant (model parameter) that controls the number of eddy events in a given time interval and $\tau$ is the eddy time scale (eddy turnover time).
This time scale is related to the total extractable (shear-available) kinetic energy which, for the instantaneous velocity vector $u_i(y,t)$ and an eddy of size $l$, is given by
\begin{equation}
  \frac{l^2}{\tau^2} \sim \frac{1}{l^4} \sum_{i=1}^{3} u_{K,i}^2 - Z \frac{\nu^2}{l^2}.
  \label{eq:eddyEnergy}
\end{equation}
Here, it follows from the construction that the kernel-weighted velocities can be summed instead of the kinetic energies.
This shows that the total extractable kinetic energy does not depend on the inter-component energy transfer, or the model parameter $\alpha$ for that matter.
Furthermore, notice the last term in equation~\eqref{eq:eddyEnergy} which represents the damping effects of the viscosity.
The model parameter $Z>0$ is used to suppress small eddy events through an energetic penalty.
This is done only to improve the numerical efficiency since very small eddy events do not contribute to the turbulent transport and are dissipated instantly by the deterministic processes.
The value $Z=1$ effectively suppresses eddy events at and below the Kolmogorov scale~\cite{Kerstein:1999}.
Values $Z>1$ have suggested for wall-bounded flows~\cite{Kerstein:1999,Schmidt_etal:2003} to account for the 3-D buffer layer dynamics (like hairpin vortices~\cite{Adrian:2007} or streaks~\cite{Lam_Banjeree:1992}) that are not resolved by ODT.

Finally, the eddy time scale $\tau$ can be computed from the instantaneous velocity vector $u_i(y,t)$ once the location $y_0$ and size $l$ of an eddy event have been selected,
\begin{equation}
  \frac{1}{\tau} = \sqrt{ \frac{1}{l^6} \sum_{i=1}^{3} u_{K,i}^2 - Z \frac{\nu^2}{l^4}}.
  \label{eq:eddyTau}
\end{equation}
This time scale is in turn compared with the mean sampling time scale $\tau_s$ to obtain the acceptance probability $p_a=\tau/\tau_s\ll1$ of a physically plausible eddy event. %% to compensate the over-sampling.
The point in time, at which equation~\eqref{eq:eddyTau} is evaluated, is obtained with the aid of a marked Poisson process.
This process assumes that eddy events are independent of each other so that the time increment between two such events can be sampled economically from an exponential distribution with the mean rate $\tau_s^{-1}$.
For further details the reader is deferred to~\cite{Kerstein:1999}

It is sometimes important to suppress unphysically large eddy events, which may occur rarely in the sampling procedure.
A large-eddy suppression (LS) mechanism is often used for this purpose (e.g.~\cite{Echekki_etal:2001}).
For confined flows, like channel flows, a simple suppression, based on the fraction of the domain length, may be sufficient.
For free shear flows, such as jets, the ``elapsed time'' method is preferred.
Only eddy events satisfying $\beta_{LS}\,\tau\leq t$ are allowed, where $\beta_{LS}$ is an additional ODT model parameter and $t$ is the current simulation time. 
Only the latter method has been used for the confined jet as it is a transient flow problem.

It is worth to note that, in the adaptive ODT formulation~\cite{Lignell_etal:2013}, the numerical resolution automatically increases in regions where eddy events occur.
This is good, since no additional adaption criteria are needed.
But, to be numerically efficient, the grid also has to be coarsened once in a while.
This is done here after some multiples of the diffusive time scale and controlled by the numerical diffusive advancement (DA) parameter $\beta_{DA}$.
For $Sc\lesssim1$, $\beta_{DA}\simeq10$ is appropriate so that coarsening takes place latest every ten viscous time units.
For $Sc\gg1$, $\beta_{DA}$ had to be increased to avoid excessive coarsening on the viscous time scale since the time step is now limited by the scalar diffusion.
For $Sc=1250$, any value $\beta_{DA}\gtrsim500$ yielded grid-independent results. 

%%%%%%%%%%%%%%%%%%%%%%%
\section{Application of ODT to the confined jet} \label{sec:config}
%%%%%%%%%%%%%%%%%%%%%%%

In the following, we describe the available reference data and discuss important aspects concerning the application of ODT to the confined jet.
These points are relevant for the interpretation of the results following below in section~\ref{sec:results}.

%%%%%%%%%%%%%%%%%%%%%%%
\subsection{Overview of the reference cases} \label{sec:ref}
%%%%%%%%%%%%%%%%%%%%%%%

Different sets of reference data are available for the confined jet.
These have been obtained with RANS~\cite{Feng2005} and LES~\cite{Kong2012}, which utilize conventional gradient-diffusion models for the subgrid-scale closure; LES-LEM~\cite{Arshad_etal:2018}, which uses the map-based stochastic linear-eddy model for this task; and laboratory measurements~\cite{Feng2005, Kong2012, Arshad_etal:2018}.
In the scope of this work, we compare ODT results with these reference data to validate the momentum transport of the model for the two Reynolds numbers $Re=20\,000$ and $40\,000$.
ODT's predictive capabilities are used to investigate the turbulent mixing of a passive scalar with Schmidt number $Sc=1$ and $1250$ for the higher Reynolds number.

The LES of Kong et al.~\cite{Kong2012} and the LES-LEM of Arshad et al.~\cite{Arshad_etal:2018} are both for the Reynolds number $20\,000$ (case~A in table~\ref{tab:Allocation_reference}).
While Kong et al.~only report the velocity statistics, Arshad et al.~provide consistent data for both the velocity and a high-$Sc$ passive scalar.
Unfortunately, the Schmidt number is very high ($Sc=2420$), while the resolution of the numerical grid and the laboratory measurements is not better than about twice the Kolmogorov length scale.
Feng et al.~\cite{Feng2005} report 2-D RANS results for $Re=40\,000$ and a passive scalar with a more moderate Schmidt number ($Sc=1250$, case~B in table~\ref{tab:Allocation_reference}). 
Even though each one of the LES approaches is much more costly than the RANS, the low-order statistical quantities exhibit surprisingly good agreement with the measured data already in the RANS.
The reason for this is not obvious but it is likely related to implicit filtering of the scalar fluctuations by not resolving the flow down to the Batchelor scale.

These implicit filtering effects concern both measurements and simulations. 
In the former, the effects manifest themselves in post-processing statistics, but for the latter they affect the flow physics captured.
In the present application, however, there are no additional physics (like chemical reactions or buoyancy) that would affect the unresolved and resolved dynamics of the reference LES.
The main effect of the small scales is therefore an increase of the dissipation and this is captured by the standard small-scale closures of the reference simulations.
So, for low-order statistical moments, it is of interest to compare results from more advanced and expensive numerical methods with those from a less expensive one, in this case the RANS results of Feng et al.~\cite{Feng2005}.
Nevertheless, we acknowledge the higher quality of the velocity statistics from the reference LES~\cite{Kong2012,Arshad_etal:2018}, which is used in the following to validate the momentum transport in ODT.

\begin{table}[t]
  \caption{%
    Summary of the cases investigated.
    $\langle N_y\rangle$ gives the typical average number of grid cells in the adaptive ODT simulations permissively resolving the Kolmogorov and Batchelor length scales.
    The near-wall resolution $y^+$ is given for comparison and only meaningful for the momentum (velocity) transport.
    The main ODT parameters $C$, $Z$, $\alpha$, the large-eddy suppression parameter $\beta_{LS}$, and the numerical parameter $\beta_{DA}$ are explained in section~\ref{sec:model}.
    All ODT simulations were conducted on local workstations with Intel$^\text{\textregistered}$ Xeon$^\text{\textregistered}$ $2.40\,\text{GHz}$ CPUs.
  }
  \centering
  \begin{tabular}{l c c c}
    \hline
    Case name & case A & \multicolumn{2}{c}{case B} \\
    \hline
    Reference & Kong et al. (2012)~\cite{Kong2012} & \multicolumn{2}{c}{Feng et al. (2005)~\cite{Feng2005}} \\
    Methods & 3-D LES, PIV & \multicolumn{2}{c}{2-D RANS, PIV/PLIF} \\
    \hline
    Fluid & water & \multicolumn{2}{c}{water with fluorescent dye} \\
    $\rho$ [kg/m$^3$] &  $1000$ &  \multicolumn{2}{c}{$1000$} \\
    $\nu$ [m$^2$/s] & $8\times10^{-7}$ & \multicolumn{2}{c}{$10^{-6}$} \\
    $U_B$ [m/s] & $0.26\bar{6}$ & \multicolumn{2}{c}{$0.66\bar{6}$} \\
    $U_0$; $U_1$ [m/s] & $0.2$; $0.4$ & \multicolumn{2}{c}{$0.5$; $1.0$} \\
    $d$ [m] & $0.02$ & \multicolumn{2}{c}{$0.02$} \\
    $x/d$ & 1; 4.5; 7.5; 12 & \multicolumn{2}{c}{1; 4.5; 7.5; 15} \\
    $Re$ & $20\,000$ & \multicolumn{2}{c}{$40\,000$} \\
    \hline
    $Sc$ & --- & $1$ & $1250$ \\
    \hline
    $C$ & $7$ & $7$ & $7$ \\
    $Z$ & $400$ & $400$ & $400$ \\
    $\alpha$ & $2/3$ & $2/3$ & $2/3$ \\
    $\beta_{LS}$ & $0.4$ & $0.4$ & $0.4$ \\
    $\beta_{DA}$ & $10$ & $10$ & $500$ \\
    $y^+$ (ODT; ref.) & $0.4$; $8$ & $0.4$; --- & $0.5$; --- \\
    $\langle N_{y} \rangle$ & $500$ & $1600$ & $2500$ \\
    CPU-h/realization & $0.004$ & $0.040$ & $5.77$ \\
    CPU-h/ensemble & $20$ & $220$ & $28\,870$ \\
    \hline
  \end{tabular}
  \label{tab:Allocation_reference}
\end{table}

%%%%%%%%%%%%%%%%%%%%%%
\subsection{ODT simulation set-up} \label{sec:bc}
%%%%%%%%%%%%%%%%%%%%%%%

The confined jet is sketched in figure~\ref{img:scheme_reactor} and has been described above.
For the stand-alone application of ODT we idealize the confined jet and consider an infinite spanwise dimension (i.e.~infinite aspect ratio).
This seems permissible since the incoming flow is due to three narrow ducts (aspect ratio~$5$ \cite{Feng2005, Kong2012, Arshad_etal:2018}).
We therefore assume that the break-up of the confined jet is not notably affected by secondary flows that are localized in the corners of the inlet ducts (e.g.~\cite{Vinuesa_etal:2018}).
In the following, we describe relevant details of the application of ODT to the confined jet.

The ODT computational domain (ODT line in figure~\ref{img:scheme_reactor}) is taken to drift with the bulk velocity $U_B$ from the inlet ($x=0$) to the end of the simulated fluid volume ($x=L$) in the course of a simulation run.
At the top and bottom boundaries ($y=\pm H/2$) zero-gradient (Neumann) conditions are prescribed for the scalar.
This is done assuming that there is no scalar flux to or from the wall.
The scalar is, thus, entirely specified by the inflow conditions. 
This is different for the velocity (momentum) for which no-slip boundary conditions are prescribed at the channel wall.

A source term is required to maintain a constant mass flux through the channel.
This is realized as a fluctuating mean pressure gradient that maintains the bulk velocity
\begin{equation}
 U_B = \dfrac{1}{H} \int_{-H/2}^{H/2} u(y) \,\mathrm{d}y = \dfrac{2U_0 + U_1}{3}, %% RX %% = \dfrac{4}{3}U_0.
 \label{eq:ubulk}
\end{equation}
where $U_0$ and $U_1$ are the bulk velocities of the upper and lower co-flow and the central jet, respectively (see table~\ref{tab:Allocation_reference}). 
Equation~\eqref{eq:ubulk} corresponds to an integral constraint on the streamwise mass flux similar to that used by Echekki et al.~\cite{Echekki_etal:2001}. % in addition to the conservative ODT model formulation (see above).
In the current implementation, a uniform but temporally fluctuating force $f_1(t)$ is computed from the developing flow field.
The source term $f_1(t)$ is computed in the deterministic advancement of equation~\eqref{eq:gov1} as a one-step correction, and exactly balances the viscous losses of the streamwise velocity.
There are no sources for the wall-normal and spanwise velocity components ($f_2=f_3=0$).

A global Galilean transformation is used to relate the elapsed simulation time~$t$ to a streamwise location~$x$.
This transformation is done for the ODT line as a whole using~\cite{Echekki_etal:2001}
\begin{equation}
 x(t) = U_B \, t .
 \label{eq:t2x}
\end{equation}
The end time $t_{end}$ of an ODT realization is, thus, implicitly defined by the length $L$ of the channel to be simulated.
Here, values of $L=12d$ and $15d$ have been used.

%%%%%%%%%%%%%%%%%%%%%%%
\subsection{Estimation of ODT model parameters} \label{sec:params}
%%%%%%%%%%%%%%%%%%%%%%%

For ODT simulations of the confined jet, the physical model parameters ($C$, $Z$, $\alpha$), the large-eddy suppression parameter ($\beta_{LS}$), and the numerical parameters (like $\beta_{DA}$) introduced in section~\ref{sec:model} had to be estimated.
This was done by ``pre-simulations'' without a passive scalar (see table~\ref{tab:Allocation_reference}) focusing on the flow properties at the downstream locations $x/d=1$ and $4.5$.
These simulations revealed that $C$ and $\beta_{LS}$ have the largest effect on the transient jet, whereas $Z$ and $\alpha$ are less important.
However, since the confined jet converges to a turbulent channel flow, for which $Z$ is not independent of $C$ (e.g.~\cite{Klein_Schmidt:2017}), we selected $Z$ consistently. 
Here, $C=7$ and $Z=400$ yields a reasonable representation of both the transient and the statistically stationary states for the two Reynolds numbers investigated.
The model parameter $\alpha$ controls the kinetic energy contained in the wall-normal and spanwise velocity components but it is generally less important for the streamwise velocity of the confined jet. 
This is due to the applied forcing mechanism (variable momentum source $f_1$), which maintains a constant mass flux regardless of the losses.

At last, note that the smallest scales are not restricted to the near-wall region even though it is common to address the grid quality with respect to $y^+=\Delta y_w \, u_\tau/\nu$, where $\Delta y_w$ is the size of the cell next to the wall and $u_\tau=\sqrt{\nu\,(\partial \langle u\rangle/\partial y)_w}$ the friction velocity.
Here, $\Delta y_w$, or $y^+$ for that matter, is a representative (average) value for the ensemble of dynamically adaptive ODT simulations.
In these simulations, the momentum boundary layer has always been well-resolved by fulfilling $y^+<1$ at various downstream locations.
In table~\ref{tab:Allocation_reference}, $y^+\approx0.5$ is reported for the representative downstream location $x/d=1$, which has been selected for comparison to the reference LES of Kong et al.~\cite{Kong2012}.

%%%%%%%%%%%%%%%%%%%%%%%
\subsection{Remarks on the computational efficiency of ODT} \label{sec:costs}
%%%%%%%%%%%%%%%%%%%%%%%

The ``pre-simulations'' mentioned above used to estimation of model parameters suggest that about $N=1000$ ensemble members are necessary to obtain a reasonable estimate of the mean velocity or the mean concentration of a passive scalar for $Sc=1$.
The root-mean-square velocity fluctuations are reasonably well estimated with at least $N=3000$ ensemble members.
The high-Schmidt-number scalar ($Sc=1250$) demands at least $N=5000$ ensemble members to converge the variance of the concentration fluctuations reasonably well since spatial scales by a factor $\approx35$ smaller than the Kolmogorov scale may occur.
This raises the question about computational efficiency.
The computational costs for the ODT simulations are given in table~\ref{tab:Allocation_reference} for both a single realization and $N=5000$ members.

The small-scale resolving ODT simulations for $Re=40\,000$, $Sc=1250$ (case~B) are roughly $100$ times slower than those for $Sc=1$.
The latter are also a factor $10$ slower than ODT simulations for $Re=20\,000$ without a passive scalar (case~A).
Assuming that one could run all members in parallel, no more than $\approx5.8\,\mathrm{h}$ are needed for $Sc=1250$. 
Similarly, the simulations for $Sc=1$ can be conducted within minutes. 
A comparative turbulent channel flow LES was performed with OpenFOAM$^\text{\textregistered}$~5.0 using the reference set-up of case~A ($N_x\times N_y\times N_z = 240\times90\times110$; Smagorinsky model~\cite{Kong2012}) using the same machine as for the ODT simulations.
This LES took $\approx5.9$ hours for ten flow through times using $15$ MPI tasks which corresponds to a total of $\approx89\,\text{CPU-h}$.
This demonstrates the capabilities of ODT to simulate canonical flows, like the confined jet, on all relevant scales using much less computational resources than conventional 3-D approaches. 

%%%%%%%%%%%%%%%%%%%%%%%
\section{Preparation of turbulent inflow conditions} \label{sec:ic}
%%%%%%%%%%%%%%%%%%%%%%%

The bulk velocity $U_B$ in equation~\eqref{eq:ubulk} and the incoming turbulence are prescribed by turbulent inflow conditions.
Using ODT, we only require an ensemble of initial flow profiles but this has to be consistent with the reference laboratory measurements~\cite{Feng2005, Kong2012, Arshad_etal:2018}.
In the following, we describe the procedure used, which circumvents the direct simulation of the entrance section ($x<0$ in figure~\ref{img:scheme_reactor}).

Flow statistics for the inflow plane are available for the present application.
Turbulent profiles are generated by combining the random fluctuation method of Lund et al.~\cite{Lund_etal:1998} with the multi-scale filtering method of Klein et al.~\cite{Klein_etal_JCP:2003}.
In practice, this involves the following steps:
\begin{enumerate}
  \item Generation of uncorrelated (pseudo-)random data from white noise.
  \item Application of the multi-scale filtering method~\cite{Klein_etal_JCP:2003} to the uncorrelated data.
  This yields the correlated random vector field $\widetilde{U}_i$, which satisfies $\langle \widetilde{U}_i \widetilde{U}_j \rangle=0$ for $i\neq j$.
  This is appropriate for homogeneous isotropic turbulence, but needs to be processed further for an application to inhomogeneous turbulence, which is the case for the confined jet.
  \item Application of the random fluctuation method~\cite{Lund_etal:1998} to the correlated data  $\widetilde{U}_i$.
  This introduces inhomogeneity by matching the prescribed (measured) mean flow and Reynolds stress tensor.
  The transformation is given by
  \begin{equation}
    u_i = \langle u_i \rangle + a_{ij} \widetilde{U}_j,
    \label{eq:u-random}
  \end{equation}
  where $\langle u_i\rangle$ is the prescribed mean velocity vector and $a_{ij}$ the prescribed correlation tensor.
  This tensor is related to the Reynolds stress tensor $R_{ij}=\langle u_i'u_j'\rangle$ and given by~\cite{Lund_etal:1998}
  \begin{equation}
    \left( a_{ij} \right) = 
    \left(
    \begin{array}{ccc}
      (R_{11})^{1/2} & 0 & 0 \\
      R_{21}/a_{11} & (R_{22} - a_{21}^2)^{1/2} & 0 \\
      R_{31}/a_{11} & (R_{32} - a_{21} a_{31})/a_{22} & (R_{33} - a_{31}^2 - a_{32}^2 )^{1/2} \\
    \end{array}
    \right).
    \label{eq:corr-tensor}
  \end{equation}
\end{enumerate}

For the confined jet, $\langle u_i\rangle$ and $R_{ij}$ are known from measurements~\cite{Feng2005, Kong2012} just behind the splitter plates.
The spanwise components were not measured so that we assume $\langle w'w'\rangle = \langle v'v'\rangle$ analogous to Kong et al.~\cite{Kong2012}. 

\begin{figure}[p]
  \centering
  \includegraphics[scale=0.5]{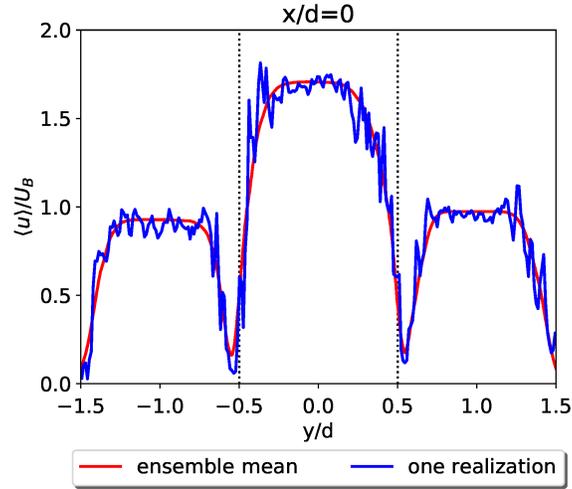}
  \caption{
    Synthetic inflow condition for case~A in terms of the streamwise velocity.
    The mean streamwise velocity profile $\langle u(y,0)\rangle$ is shown together with a single realization of a streamwise velocity profile $u(y,0)$.
    Dotted lines give the locations of the splitter plates.
    The reference data (PIV, LES) is from~\cite{Kong2012}.
  }
  \label{img:Comparison_synthetic_initial_condition_Kong}
\end{figure}

\begin{figure}[p]
  \centering
  \includegraphics[scale=0.5]{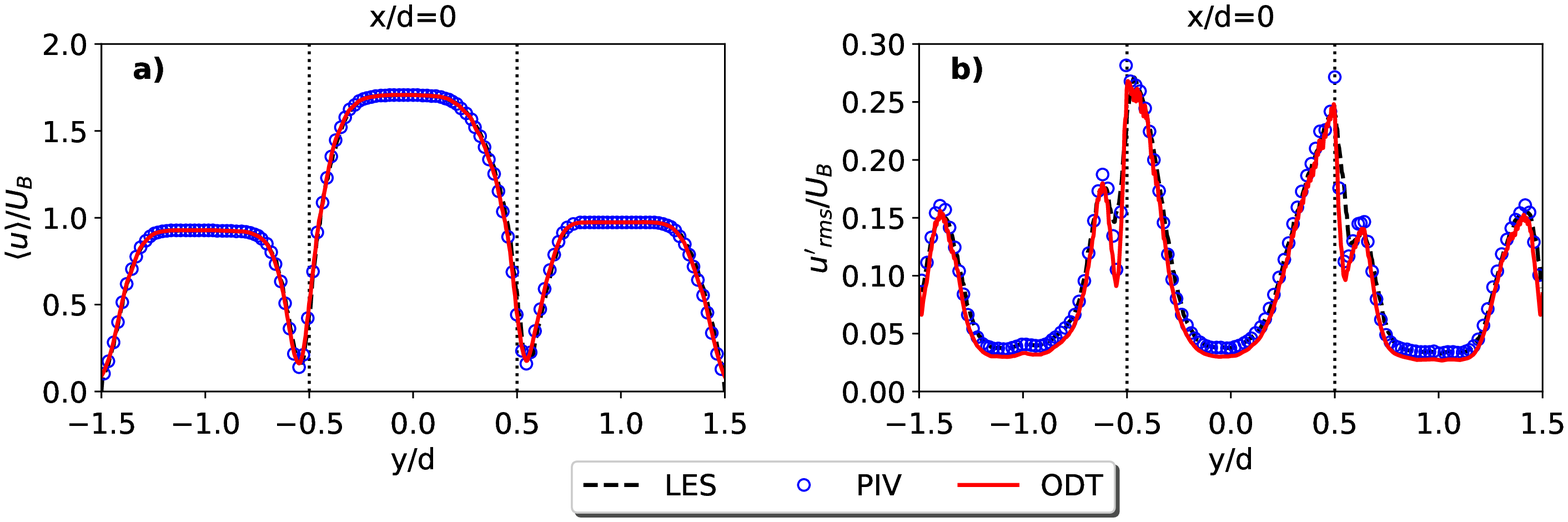}
  \caption{
    Mean streamwise velocity (a) and r.m.s.~streamwise velocity fluctuations (b) for case~A comparing present synthetic inflow conditions with reference data (PIV, LES) from~\cite{Kong2012}.
    All curves collapse very well on each other.
  }
  \label{img:synthetic_initial_condition_Kong}
\end{figure}

\begin{figure}[p]
  \centering
  \includegraphics[scale=0.5]{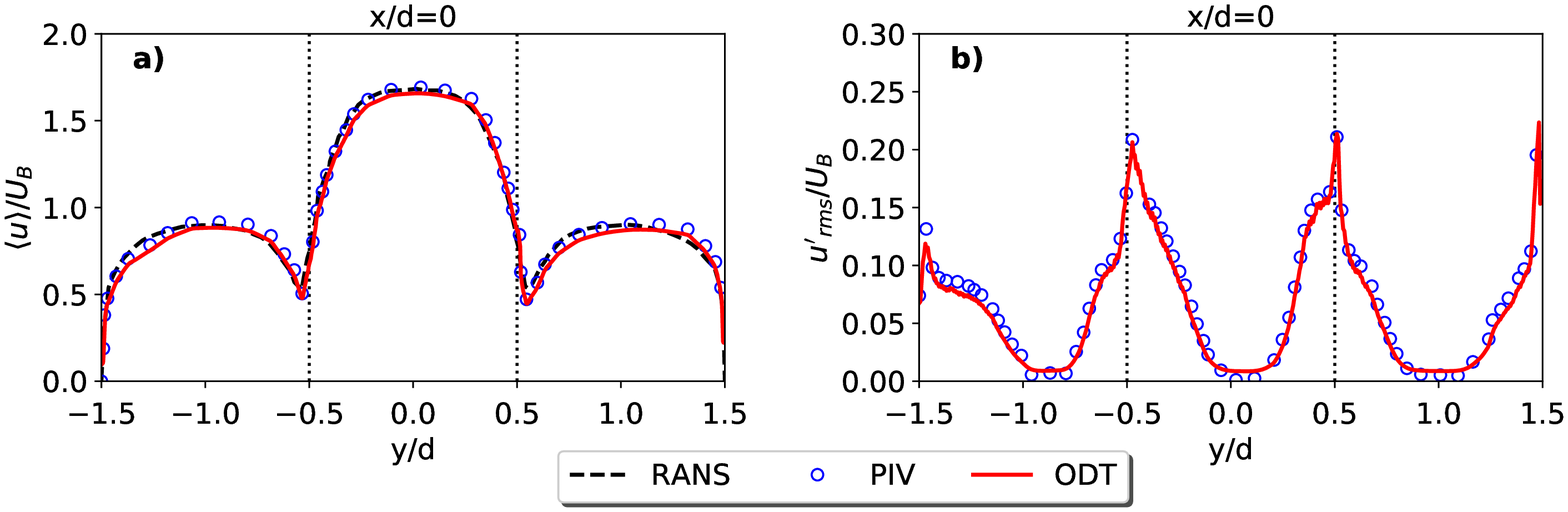}
  \caption{
    Mean streamwise velocity (a) and r.m.s.~streamwise velocity fluctuations (b) for case~B comparing present synthetic inflow conditions with reference data (PIV, RANS) from~\cite{Feng2005}.
    All curves collapse very well on each other.
  }
  \label{img:synthetic_initial_condition_Feng}
\end{figure}

Figure~\ref{img:Comparison_synthetic_initial_condition_Kong} shows the ensemble-averaged initial streamwise velocity profile $\langle u(y,0)\rangle$ together with a single synthetic profile $u(y,0)$ for case~A.
The fluctuations occur on different length scales and they are largest just behind the splitter plates, which are located at $y/d=\pm0.5$ as indicated by dotted vertical lines.
In addition, figures~\ref{img:synthetic_initial_condition_Kong} and \ref{img:synthetic_initial_condition_Feng} show the corresponding mean velocity and root-mean-square (r.m.s.) velocity fluctuations for case~A and B, respectively.
In both cases, the synthetic initial conditions are statistically consistent with the reference measurements and initial conditions of the reference LES.
This is an important property, which cannot be obtained with ODT as turbulent inflow generator (see~\ref{sec:ic-odt}).

%%%%%%%%%%%%%%%%%%%%%%%
\section{Results} \label{sec:results}
%%%%%%%%%%%%%%%%%%%%%%%

In the first three subsections, the turbulent mixing of the momentum is addressed by looking into the velocity statistics of the constant-density flow according to case~A (see table~\ref{tab:Allocation_reference}).
In the remaining subsections, the turbulent mixing of a passive scalar is investigated using the set-up of case~B.
The comparison between the ODT results and the available reference data is made at four different locations downstream of the inflow plane (see figure~\ref{img:scheme_reactor}).
For all cases shown, an ensemble of $N=5000$ members (ODT realizations) has been simulated and statistically analyzed as described in \ref{sec:stats}.

%%%%%%%%%%%%%%%%%%%%%%%
\subsection{Mean streamwise velocity profiles} \label{sec:umean}
%%%%%%%%%%%%%%%%%%%%%%%

\begin{figure}[tp]
  \centering
  \includegraphics[scale=0.4]{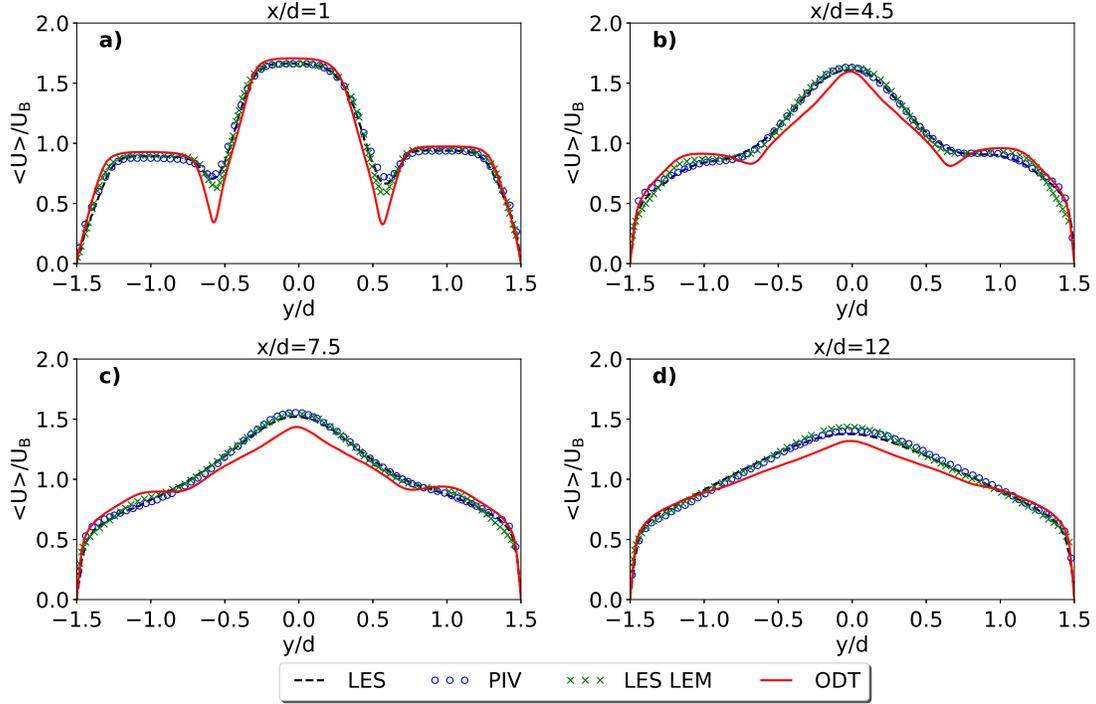}
  \caption{%
    Mean streamwise velocity $\langle u\rangle$ profiles for various downstream locations $x/d=\text{const}$.
    ODT results are shown for case~A ($Re=20\,000$) together with the corresponding reference data from PIV and LES~\cite{Kong2012}, as well as LES-LEM~\cite{Arshad_etal:2018}.
  }
  \label{img:Umean_Kong}
\end{figure}

\begin{figure}[tp]
  \centering
  \includegraphics[scale=0.4]{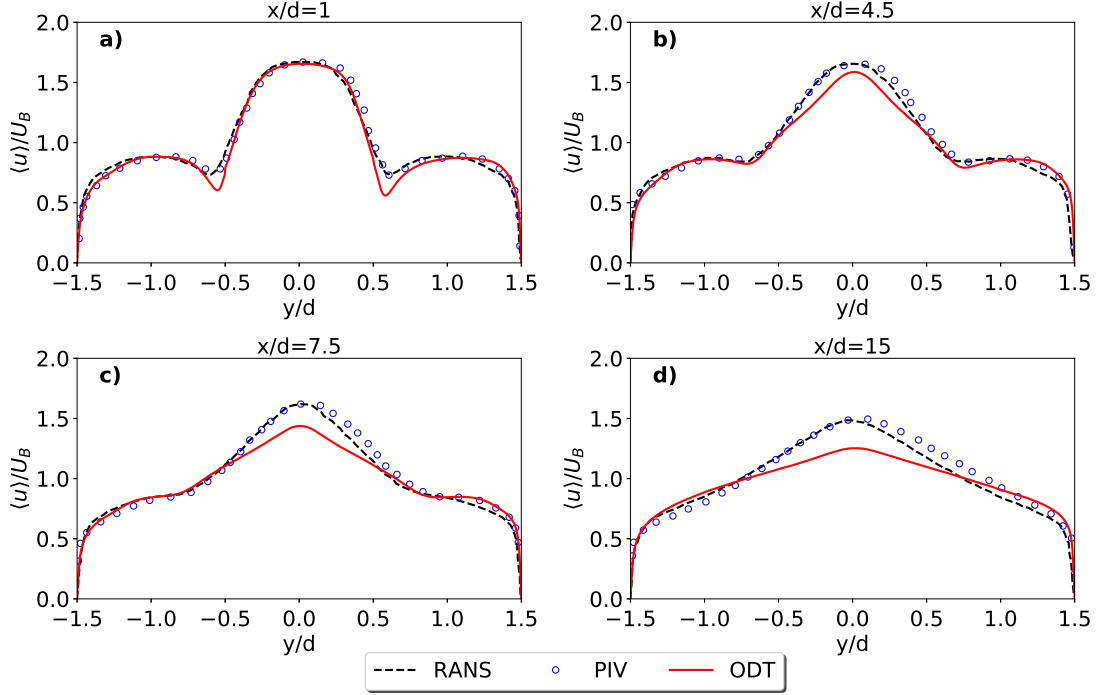}
  \caption{%
    Same as figure~\ref{img:Umean_Kong} but for case~B ($Re=40\,000$) together with the corresponding reference data from PIV and RANS~\cite{Feng2005}.
  }
  \label{img:Umean_Feng}
\end{figure}

Figures~\ref{img:Umean_Kong} and \ref{img:Umean_Feng} show profiles of the mean streamwise velocity $\langle u\rangle$ normalized with the constant bulk velocity $U_B$ at predefined downstream locations for two different Reynolds numbers. 
In both figures, the measured reference profiles are not fully symmetric.
This manifests itself by a weak skew, which is partly captured by the reference LES and ODT results due to the prescribed inflow conditions.
In figure~\ref{img:Umean_Feng} (case~B), there is an additional upwards shift of the PIV profiles, which is likely a systematic measurement error~\cite{Feng2005}.

Up to $x/d=1$, the flow is strongly influenced by the inflow condition.
In this near-field region of the inlet, the ODT large-eddy suppression mechanism (model parameter $\beta_{LS}$) is important for an accurate representation of the transient stage given by the spreading of the jet.
Here, $\beta_{LS}=0.4$ yields results that closely resemble the available reference data.
The selected value of $\beta_{LS}$ is well within the range $[0.34, 1.45]$ suggested by Echekki et al.~\cite{Echekki_etal:2001} for planar jet flames.

For both cases investigated (figures~\ref{img:Umean_Kong} and \ref{img:Umean_Feng}), the ODT results obtained at $x/d=1$ underestimate the mean velocity in the wake of the splitter plates but slightly overestimate the local maximum velocities.
This is a lower-order modeling error, because $U_B$ is kept constant in the present ODT simulations.
Further downstream, at $x/d\geq4.5$, the wake of the splitter plates differs somewhat from the reference data.
In the ODT results, the wake is notably more persistent.
Correspondingly, a more pronounced central peak can be discerned.
The maximum jet velocity, however, is weaker than in the reference data.
These are known phenomena of the ODT modeling approach and normally attributed to unresolved 3-D flow structures within the ODT modeling framework (e.g.~\cite{Kerstein_etal:2001, Lignell_etal:2013, Schulz_etal:2013, Klein_Schmidt:2017}).
Here, these flow structures are mainly related to the details of the breaking jet (see~\cite{Kong2012, Arshad_etal:2018}). 
The turbulent boundary layer flow and the duct geometry of the reference set-up might also play a role, but we estimate that this is only relevant far downstream ($x/d\gg1$), when the jet has reached the wall. 

%%%%%%%%%%%%%%%%%%%%%%%
\subsection{Root-mean-square velocity fluctuations} \label{sec:ufluc}
%%%%%%%%%%%%%%%%%%%%%%%

\begin{figure}[p]
  \centering
  \includegraphics[scale=0.4]{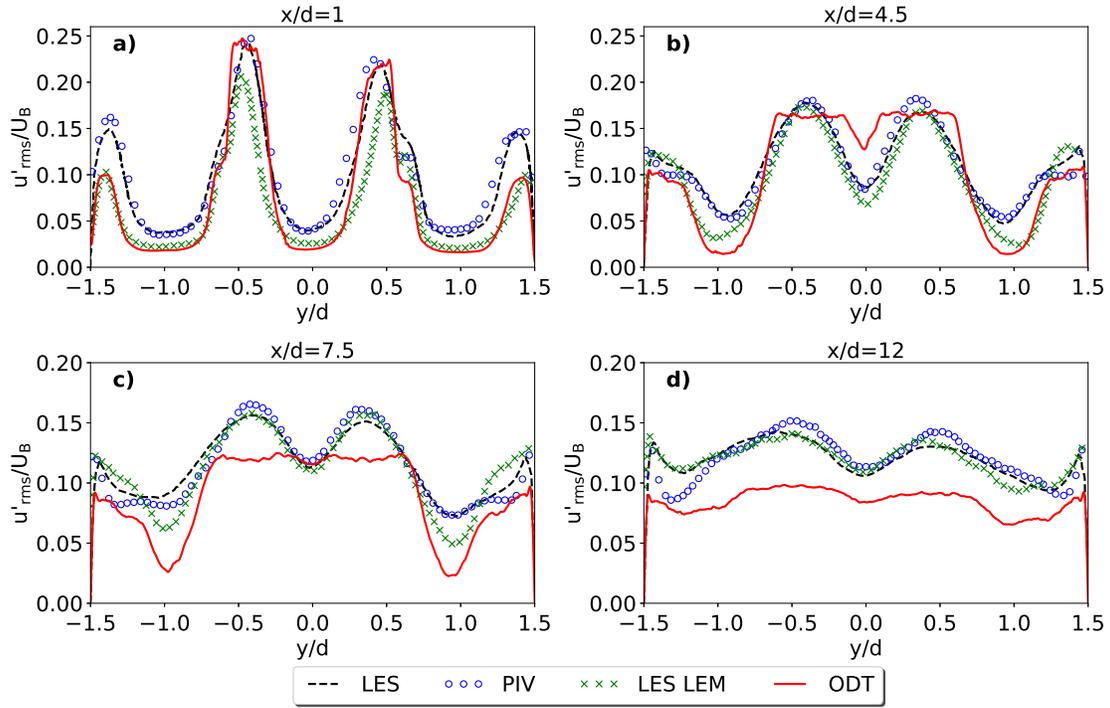}
  \caption{%
    Streamwise root-mean-square velocity $u'_{rms}$ profiles for various downstream locations $x/d=\text{const}$.
    ODT results are shown for case~A together with the corresponding reference data from PIV and LES~\cite{Kong2012}, as well as LES-LEM~\cite{Arshad_etal:2018}.
  }
  \label{img:urms_Kong}
\end{figure}

\begin{figure}[p]
  \centering
  \includegraphics[scale=0.4]{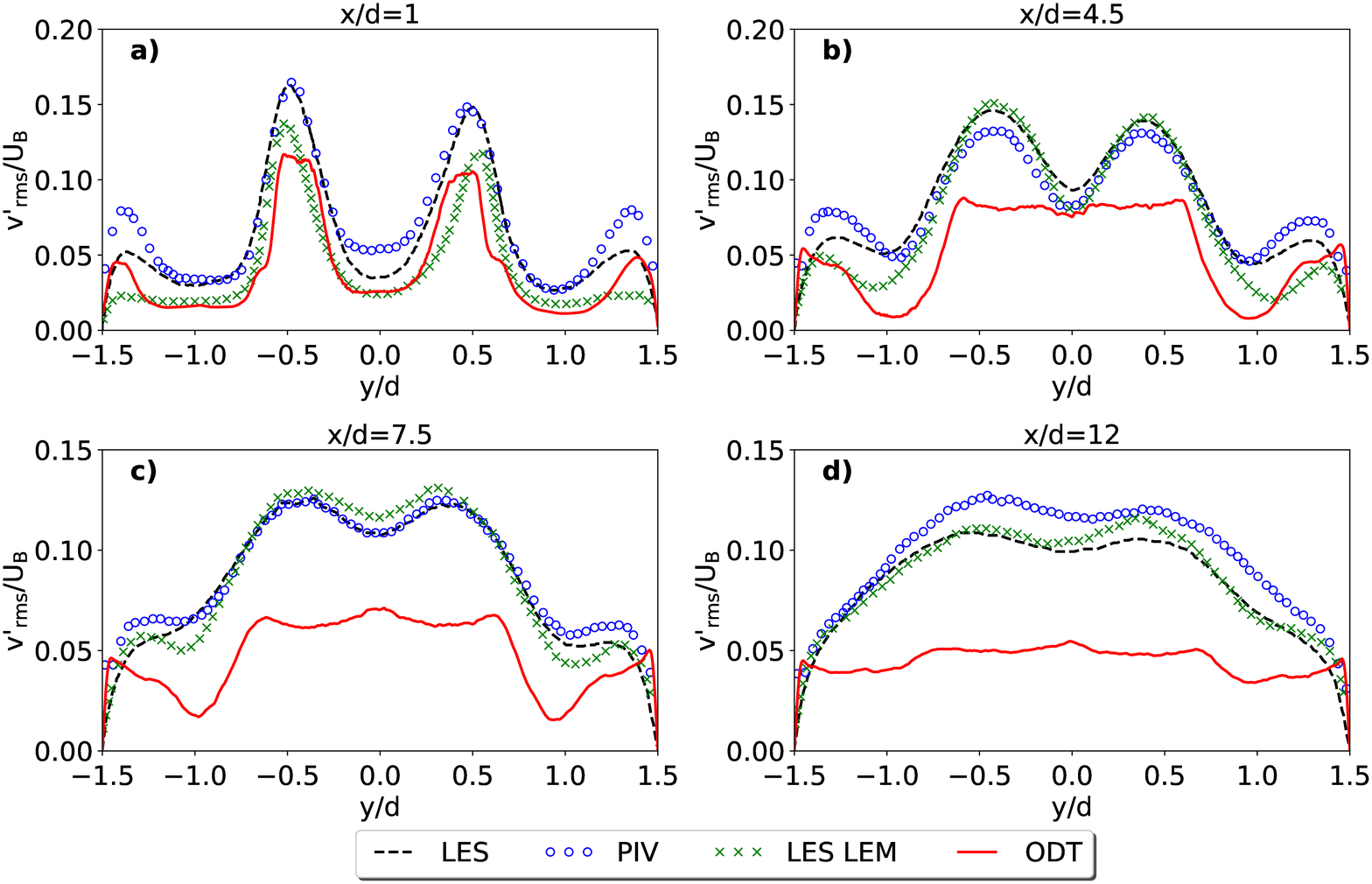}
  \caption{
    Same as figure~\ref{img:urms_Kong} but for the wall-normal root-mean-square velocity~$v'_{rms}$.
  }
  \label{img:vrms_Kong}
\end{figure}

Figure~\ref{img:urms_Kong} shows the streamwise ($u'_{rms}$) root-mean-square velocity fluctuations and figure~\ref{img:vrms_Kong} the wall-normal ($v'_{rms}$)ones for various downstream locations.
Four local maximums can be discerned that are located in the wake of the splitter plates ($y/d\approx\pm0.5$) and the turbulent boundary layer at the channel wall ($y/d\approx\pm1.5$).

Close to the inlet, at location $x/d=1$, the ODT results exhibit very good qualitative agreement with the reference LES of Kong et al.~\cite{Kong2012} and an even better one with the LES-LEM of Arshad et al.~\cite{Arshad_etal:2018}.
This indicates that, in the near-field of the inlet, both ODT and LES-LEM results are dominated by the small-scale stochastic transport.
Further downstream, at $x/d=4.5$, the ODT results diverge from the reference LES and LES-LEM results but a fair agreement is still obtained.
For $x/d\geq7.5$, in the bulk and near the wall, ODT underestimates the fluctuations $u'_{rms}$ by $\approx30\%$, which is a well-known limitation of the model (e.g.~\cite{Kerstein:1999, Kerstein_etal:2001, Lignell_etal:2013}).

The wall-normal velocity fluctuations $v'_{rms}$ (figure~\ref{img:vrms_Kong}) are very similar to the streamwise ones, except that ODT yields systematically lower values of $v'_{rms}$ and, in particular, far downstream of the inlet.
This is mainly related to the lower-order formulation of the model, in particular, to the treatment of the pressure fluctuations (see equations~\eqref{eq:eddy2} and \eqref{eq:ci}).
The model parameter $\alpha$, therefore, has a certain influence on the observed difference~\cite{Ashurst_Kerstein:2005}.
The results for $v'_{rms}$ suggest that the present ODT simulation exhibits a faster convergence to the fully-developed turbulent state than the reference LES.
In this asymptotic state both $u'_{rms}$ and $v'_{rms}$ exhibit a local minimum at the centerline $y/d=0$. 

Note that the ODT spanwise velocity fluctuations $w'_{rms}$ are presently identical to the wall-normal ones, that is, $w'_{rms}=v'_{rms}$.
This property is imposed initially by the prescribed inflow conditions (see section~\ref{sec:ic}) and is then maintained in downstream direction by the model.

%%%%%%%%%%%%%%%%%%%%%%%
\subsection{Reynolds stress} \label{sec:reystress}
%%%%%%%%%%%%%%%%%%%%%%%

\begin{figure}[t]
  \centering
  \includegraphics[scale=0.4]{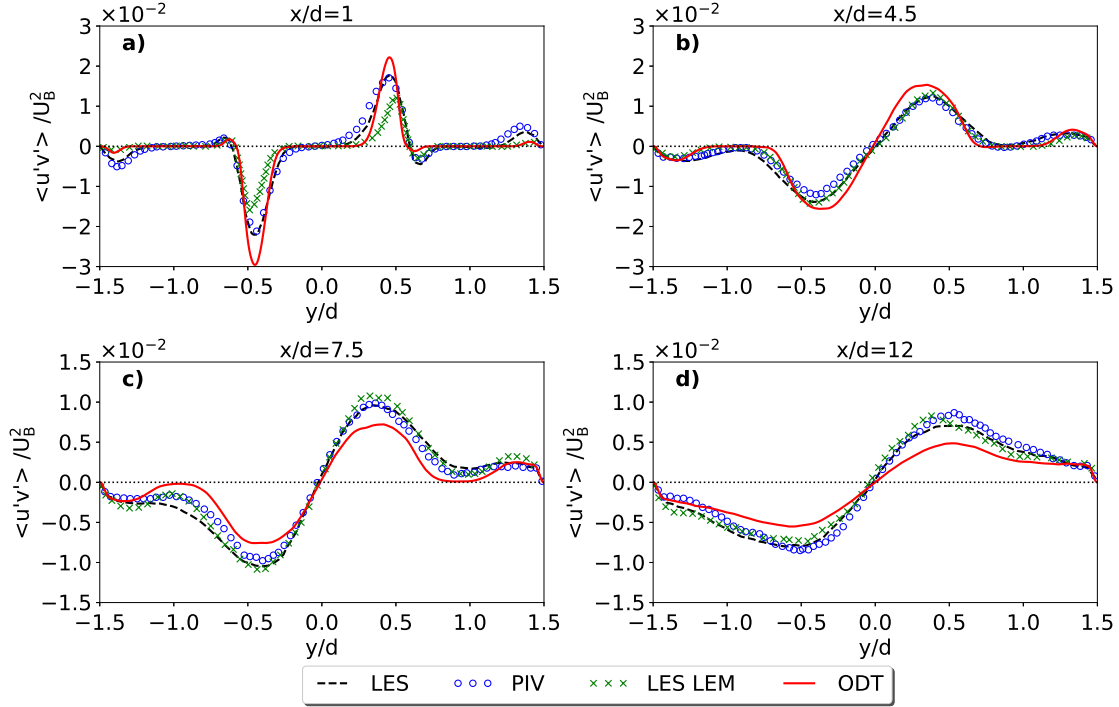}
  \caption{%
    Reynolds stress $\langle u'v'\rangle$ profiles for various downstream locations $x/d=\text{const}$.
    ODT results are shown for case~A together with the corresponding reference data from PIV and LES~\cite{Kong2012}, as well as LES-LEM~\cite{Arshad_etal:2018}.
    The black dotted line gives the zero level.
  }
  \label{img:uv_Kong}
\end{figure}

Figure~\ref{img:uv_Kong} shows wall-normal profiles of the Reynolds stress component $\langle u'v'\rangle$.
Here, and for all downstream locations, the present ODT results are in good agreement with the available reference data.
This may seem surprising at first since the fluctuations $u'_{rms}$ and $v'_{rms}$ themselves (see above) are notably different from the reference data for the downstream locations $x/d\geq4.5$.
The reason, why $\langle u'v'\rangle$ is rather well-represented in ODT, is due to the map-based advection.
As described in \ref{sec:stats}, the Reynolds stress $\langle u'v'\rangle$ is \emph{not} just the average of $u'v'$ (or, following equation~\eqref{eq:eddy2}, the average of $u''v''$).
Instead, it is given by the accumulation of eddy events.

Nevertheless, at some location the Reynolds stress $\langle u'v'\rangle$ has dropped notably below the reference values (see figure~\ref{img:uv_Kong}, $x/d\geq7.5$).
This implies reduced turbulence intensity and, thus, less mixing in ODT.
This effect seems to be the cause for the more persistent central peak and wake of the splitter plates in the ODT mean velocity profiles (figure~\ref{img:Umean_Kong}).

Furthermore, for $x/d=1$, the ODT Reynolds stresses shown in figure~\ref{img:uv_Kong} are slightly overestimating the reference data in the wake of the splitter plates ($y/d\approx\pm0.5$) but underestimating near the channel wall ($y/d\approx\pm1.5$).
Unexpectedly, the strongest differences are observed between the stochastic ODT and LES-LEM~\cite{Arshad_etal:2018} results.
For $x/d=4.5$, the overall agreement is better but, from here on, the present ODT results exhibit somewhat too much turbulent transport (e.g.~around $y/d\approx\pm0.3$).
This increased turbulence activity explains the differences seen in the mean velocity profiles as noted in the discussion of figure~\ref{img:Umean_Kong} for $x/d\geq4.5$.
Altogether, the wall-normal turbulent momentum transport given by the Reynolds stress $\langle u'v'\rangle$ is rather well-captured by ODT for all downstream locations.

%%%%%%%%%%%%%%%%%%%%%%%
\subsection{2-D visualizations of the scalar concentration} \label{sec:2d}
%%%%%%%%%%%%%%%%%%%%%%

In the following, we address the turbulent mixing of a passive scalar of a low ($Sc=1$) and high Schmidt number ($Sc=1250$), respectively.
Due to the available reference data, we switch to case~B with the Reynolds number $Re=40\,000$ (see table~\ref{tab:Allocation_reference}) for the rest of this paper. 
The case of a low Schmidt number is considered in addition to address the effect of small-scale dissipation.
As sketched in figure~\ref{img:scheme_reactor}, the passive scalar is always prescribed by a homogeneous concentration ($\phi=1$) in the small central channel but it is absent ($\phi=0$) in the outer two.
The ODT configuration is the same as above but, in order to resolve the flow down to the Batchelor scale, the grid-adaption range has been adjusted.

\begin{figure}[t]
  \centering
%   %%%% >>>
%   \ \\[4ex] !! ATTENTION: FIG 10(a) ATTACHED AS TIFF !! \\[4ex] 
%   %%%% <<<
  \includegraphics[width=0.7\linewidth]{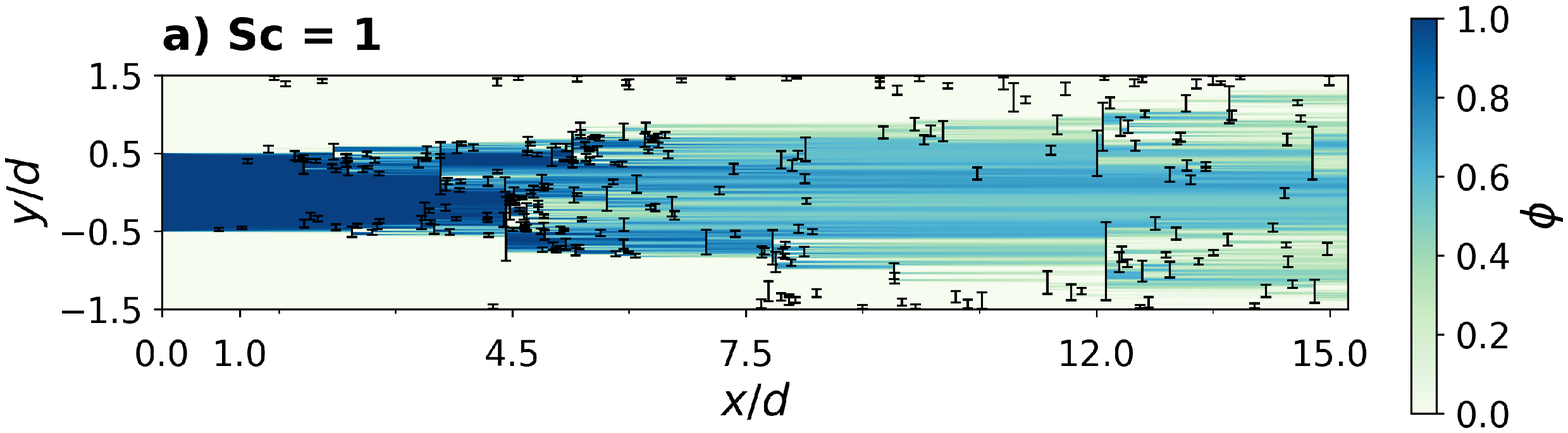}
  \includegraphics[width=0.7\linewidth]{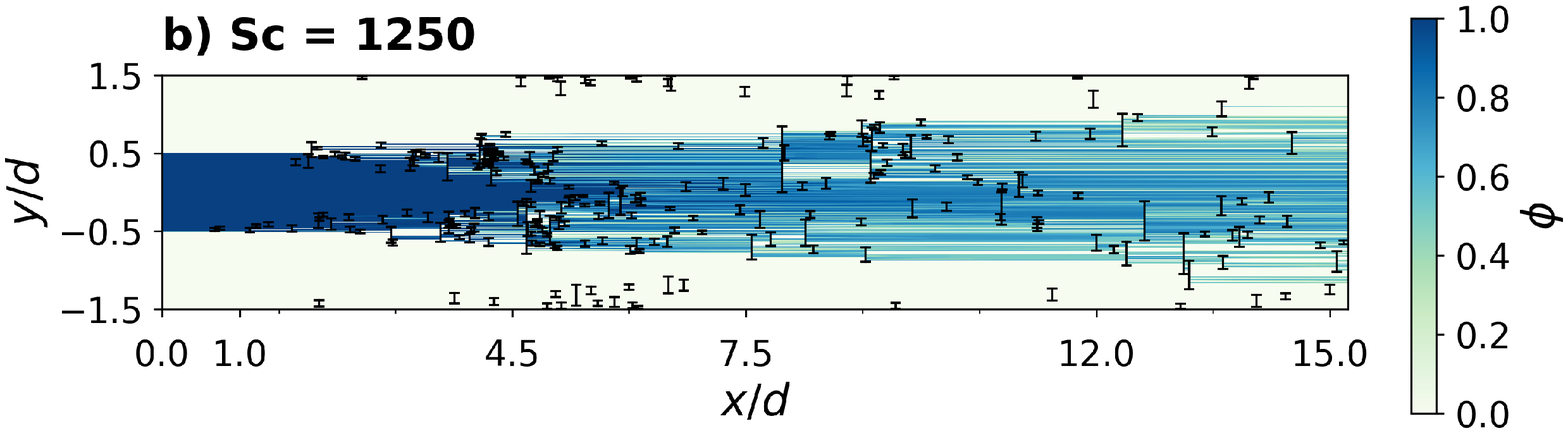}
  \caption{%
    2-D visualizations of the passive scalar concentration $\phi$ in a streamwise wall-normal plane (see figure~\ref{img:scheme_reactor}).
    Two ODT solutions are shown, one for $Sc=1$~(a) and one for $Sc=1250$~(b) using $Re=40\,000$ (case~B).
    The turbulent inflow enters on the left at $x/d=0$.
    Short black lines give the size and position of instantaneous eddy events.
    Eddy events in the sidewall boundary layer do not contribute to the scalar mixing upstream of $x/d\approx12$.
  }
  \label{img:vis2d_Feng}
\end{figure}

Figure~\ref{img:vis2d_Feng} shows 2-D snapshots of the scalar concentration $\phi(x,y)$ in the confined jet obtained with ODT for $Sc=1$ and $1250$, respectively.
These snapshots lack vortical structures compared to full 3-D numerical simulations, because of the limitations of the modeling approach, but each of them is statistically representative.
This concerns, in particular, the spatial scales in the numerical solution.

To investigate this further, let us consider a laminar jet for a moment.
This jet would broaden only due to molecular diffusion.
The average (bulk) residence time of the fluid is $t_R=L/U_B$, which is the same here as in the case of the turbulent jet due to the fixed mass flux.
The diffusive broadening can be measured by the length scales $\delta_u$ and $\delta_\phi$ for the velocity (momentum) and the passive scalar, respectively.
For case~B with $Re=40\,000$ and $Sc=1250$ (table~\ref{tab:Allocation_reference}), one obtains
\begin{equation}
  \delta_u/d\simeq\sqrt{\nu t_R}/d \simeq \sqrt{45}\, Re^{-1/2} \approx 0.03,
  \qquad
  \delta_\phi/d\simeq Sc^{-1/2} \delta_u/d \approx 0.001.
  \label{eq:visc-broadening}
\end{equation}

In the case of the turbulent jet, the scalar has reached the wall for $x/d\approx 15$.
This implies a turbulent broadening or increase of the mixing efficiency by $\approx30$ for the velocity or the $Sc=1$ scalar, and by $\approx1000$ for the $Sc=1250$ scalar.
In the mixed region, however, the scalar fluctuations are not dissipating quickly but persist for a long time and, thus, affect the fluid far downstream of the inlet.
This has implications for the resolution requirements of numerical simulations, for example, once chemical reactions or buoyancy are considered, which demand a proper representation of the spatio-temporal variability of the scalar concentration.

%%%%%%%%%%%%%%%%%%%%%%%
\subsection{Mean scalar concentration} \label{sec:sca-mean}
%%%%%%%%%%%%%%%%%%%%%%%

\begin{figure}[p]
  \centering
  \includegraphics[scale=0.4]{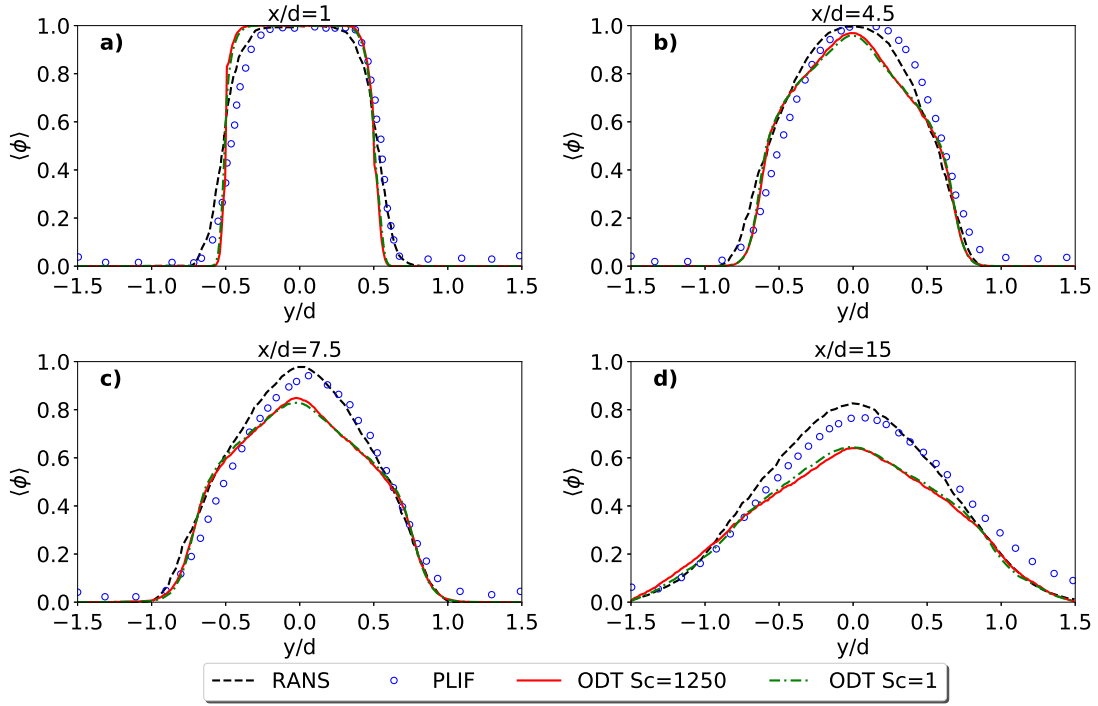}
  \caption{%
    Mean scalar concentration $\langle\phi\rangle$ for various downstream locations $x/d=\text{const}$.
    ODT results are shown for case~B ($Re=40\,000$, $Sc=1250$) together with the corresponding reference data (PLIF, RANS with nominally the same $Sc$)~\cite{Feng2005}.
    There is no reference corresponding to the ODT results with $Sc=1$. 
  }
  \label{img:Tmean_Feng}
\end{figure}

Figure~\ref{img:Tmean_Feng} shows the mean scalar concentration $\langle\phi\rangle$ for various Schmidt numbers and downstream locations.
The ODT results for $Sc=1$ and $1250$ are shown together with the under-resolved reference data for $Sc=1250$ and all of then exhibit good agreement.
It is worth to note that the ODT results for the mean field are independent of the Schmidt number since, here, the turbulent advective transport processes dominate over the molecular diffusive ones as discussed in the previous section.
Next, we will focus on the differences.

The reference scalar concentration obtained by PLIF and RANS~\cite{Feng2005} differ mainly by a small offset in wall-normal direction but they are otherwise in good agreement.
The skew of the mean scalar concentration results from the weakly asymmetric inflow profile prescribed for the simulations (see section~\ref{sec:ic}).
This skew is more notable in the ODT results.
For $x/d=1$, the ODT results exhibit a larger mean scalar concentration than the reference data, which indicates less mixing in the near field of the inflow plane.
For $x/d\geq4.5$, by contrast, the ODT results exhibit a smaller concentration. 
This effect is more convincingly revealed by table~\ref{tab:Conservation_scalar}, which gives the integrated mean scalar concentration $\phi_B$ for the selected downstream locations $x/d$ (see figure~\ref{img:scheme_reactor}).
In the lower-order ODT formulation, $\phi_B$ is given by the integral
\begin{equation}
 \phi_B = \dfrac{1}{H} \int_{-H/2}^{H/2} \langle\phi(y)\rangle \,\mathrm{d}y.
 \label{eq:phiB}
\end{equation}
This quantity is supposed to correspond to the central plane of the large channel (duct) of the references.

\begin{table}[p]
    \caption{%
      Normalized mean scalar concentration $(H/d)\,\phi_B$ according to equation~\eqref{eq:phiB} at various downstream locations $x/d=\text{const}$ for case~B ($Re=40\,000$, $Sc=1250$).
      The confidence interval of the reference data has been estimated from the scalar variance of~\cite{Feng2005}.
      Values are accurate to the discretization error ($\lesssim10^{-5}$) when no margin is given.
      ODT conserves the scalar for the 1-D computational domain, which is different compared to the central plane of the available reference data.
    }
    \centering
    \begin{tabular}{c c c c}
        \hline
        $x/d$ & RANS~\cite{Feng2005} & PLIF~\cite{Feng2005} & ODT \\
        \hline
         0.0 & 1.0 & 1.0 & 1.0 \\
         1.0 & $1.030\pm0.002$ & $1.04\pm0.13$ & 1.0 \\
         4.5 & $1.074\pm0.001$ & $1.11\pm0.09$ & 1.0 \\
         7.5 & $1.12\pm0.02$   & $1.10\pm0.08$ & 1.0 \\
        15.0 & $1.21\pm0.05$   & $1.21\pm0.06$ & 1.0 \\
        \hline
    \end{tabular}
    \label{tab:Conservation_scalar}
\end{table}

The passive scalar is conserved by the ODT formulation for the lower-order computational domain (ODT line).
The reference measurements and RANS simulations of Feng et al.~\cite{Feng2005}, as well as the LES-LEM of Arshad et al.~\cite{Arshad_etal:2018}, however, exhibit an accumulation of the scalar in the central plane of the channel.
The reason for this difference is likely related to secondary flows in the reference duct geometry, which is not resolved by ODT.
In the duct, corner vortices~\cite{Vinuesa_etal:2018} are present and likely cause a temporary accumulation of the scalar in the center of the 3-D flow domain.
However, it remains open why the 2-D RANS simulation of Feng et al.~\cite{Feng2005} (2-D channel with modeled small-scale 3-D turbulence) captures the downstream increase of the scalar for the central plane of the channel.

Altogether, the mean scalar distribution (figure~\ref{img:Tmean_Feng}) obtained with the small-scale resolving ODT simulations is very similar to that of the coarse-resolution reference RANS simulations and measurements.
Hence, if one is \emph{only} interested in the mean values, no substantial improvement can be expected from more expensive methods (like LES or DNS), nor the small-scale resolving ODT simulations.

%%%%%%%%%%%%%%%%%%%%%%%
\subsection{Scalar fluctuation variance} \label{sec:sca-var}
%%%%%%%%%%%%%%%%%%%%%%%

It was shown above that the velocity fluctuations, $u'_{rms}$ and $v'_{rms}$, are typically underestimated by $\approx30$--$50\%$ in the ODT results even though the flow is resolved on all scales.
This serves as a reference for the passive scalar for which, however, we will next see a different behavior and a sensitivity to the spatial resolution.

\begin{figure}[t]
  \centering
  \includegraphics[scale=0.4]{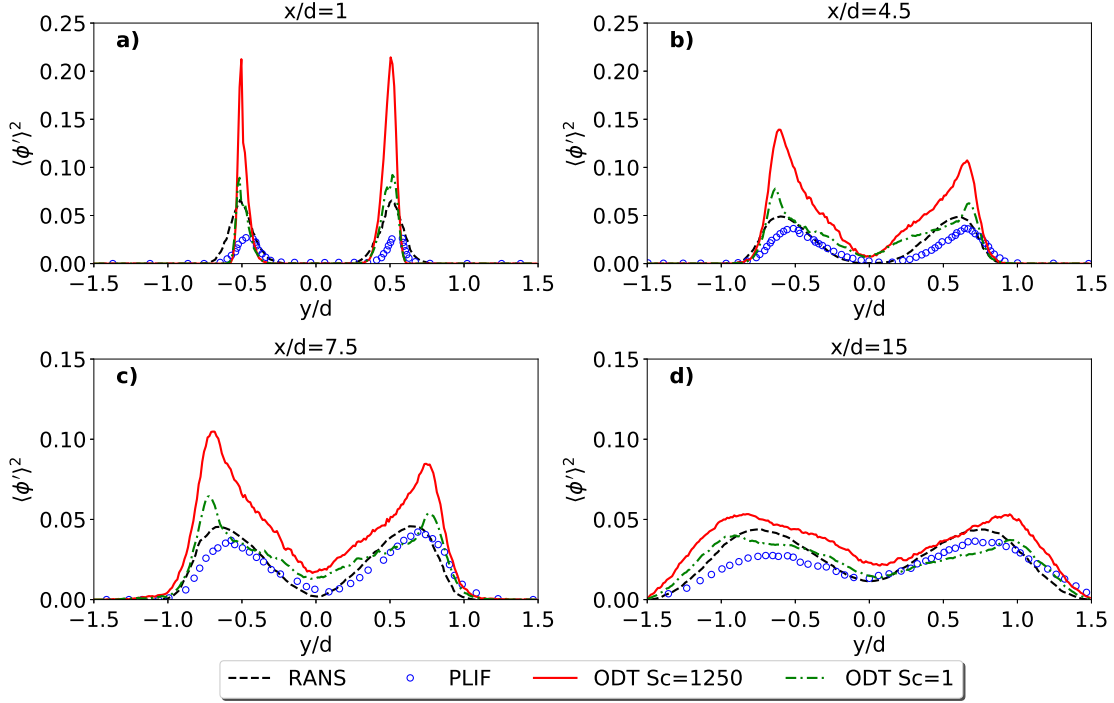}
  \caption{%
    Same as figure~\ref{img:Tmean_Feng} but for the scalar variance $\langle \phi^{\prime\,2}\rangle$.
  }
  \label{img:Trms_Feng}
\end{figure}

Figure~\ref{img:Trms_Feng} shows the scalar variance $\langle\phi^{\prime\,2}\rangle$ for various Schmidt numbers and downstream locations corresponding to case~B.
All pointwise fluctuation statistics shown exhibit a bimodal structure with local maximums in the wake of the splitter plates.
These qualitative features are well-captured by the ODT model in comparison to the reference data~\cite{Feng2005}.
In addition, the spreading in terms of the leading edge variance is very well captured by ODT.
This indicates again that the break-up of the jet is reasonably well captured by ODT.

Substantial differences, however, are visible with respect to the magnitude of the scalar fluctuations for nominally the same Schmidt number $Sc=1250$ in ODT as well as the reference PLIF and RANS data.
Near the inlet, at $x/d=1$, the ODT results exhibit $5$--$10$ times larger values than the reference data.
Further downstream, at $x/d=15$, this factor has dropped to $\approx1.2$--$2$. 
Nevertheless, the relatively \emph{larger} scalar variance in ODT is to be seen in contrast to the relatively \emph{smaller} velocity fluctuations discussed above.
We attribute this effect to the proper small-scale resolution, which has been achieved with ODT but not in the reference data.
The scalar variance is, precisely for this reason, somewhat asymmetric in ODT due to the slightly skewed inflow data (see section~\ref{sec:ic}).

As mentioned above, the reference data of Feng et al.~\cite{Feng2005} as well as that of Arshad et al.~\cite{Arshad_etal:2018} (not shown here due to different Schmidt and Reynolds numbers), do not resolve the Batchelor scales.
Both measuring and numerical techniques exhibit an effective resolution, which is roughly comparable to $2$--$5$ Kolmogorov length scales (compare with table~\ref{tab:Allocation_reference}). 
In the present application, the scalar fluctuations are only dissipated on the small-scales without any feedback on the flow.
Therefore, in the coarse-resolution reference data, these fluctuations are subject to filtering and, in the RANS simulations, they are even prescribed by modeling assumptions (like a tailored turbulent Schmidt number).

In order to mimic the coarse resolution of the reference data, ODT simulations have been performed for case~B with $Sc=1$ to increase the small-scale dissipation.
These simulations are still well-resolved but the Batchelor scale is now equal to the Kolmogorov scale.
The results are shown in figure~\ref{img:Trms_Feng}.
Indeed, a lower Schmidt number yields significantly lower values of the scalar variance, which is especially notable close to the inlet.
By changing from $Sc=1250$ to $1$, the scalar variance reduces $\approx70\%$ for $x/d=1$, $\approx50\%$ for $4.5\leq x/d\leq 7.5$, and approaches zero for $x/d\geq15$, that is, far downstream of the inlet in the well-mixed region of the confined jet.
Altogether, the well-resolved ODT results for $Sc=1$ agree much better with the under-resolved reference data at $Sc=1250$.
This indicates that both high-resolution numerical simulations and laboratory measurements are necessary when the local variability of a scalar property is of concern.
ODT's predictive capabilities suggest that high-resolution laboratory measurements, LES or DNS should exhibit a similar behavior.
Such 3-D simulations would be costly but high-resolution LES-LEM~\cite{Arshad_etal:2018} or ODTLES~\cite{Glawe_etal:2018} seem to be in reach for $Sc\lesssim100$.

%%%%%%%%%%%%%%%%%%%%%%%
\subsection{Turbulent scalar fluxes} \label{sec:sca-flux}
%%%%%%%%%%%%%%%%%%%%%%%

The turbulent scalar fluxes are the result of an ensemble of ODT simulations.
The streamwise component $\langle u'\phi'\rangle$ is computed straightforwardly for each location.
The wall-normal component $\langle v'\phi'\rangle$, however, is computed by the map-induced transport analogous to the Reynolds stress as described in \ref{sec:stats}.

\begin{figure}[p]
  \centering
  \includegraphics[scale=0.4]{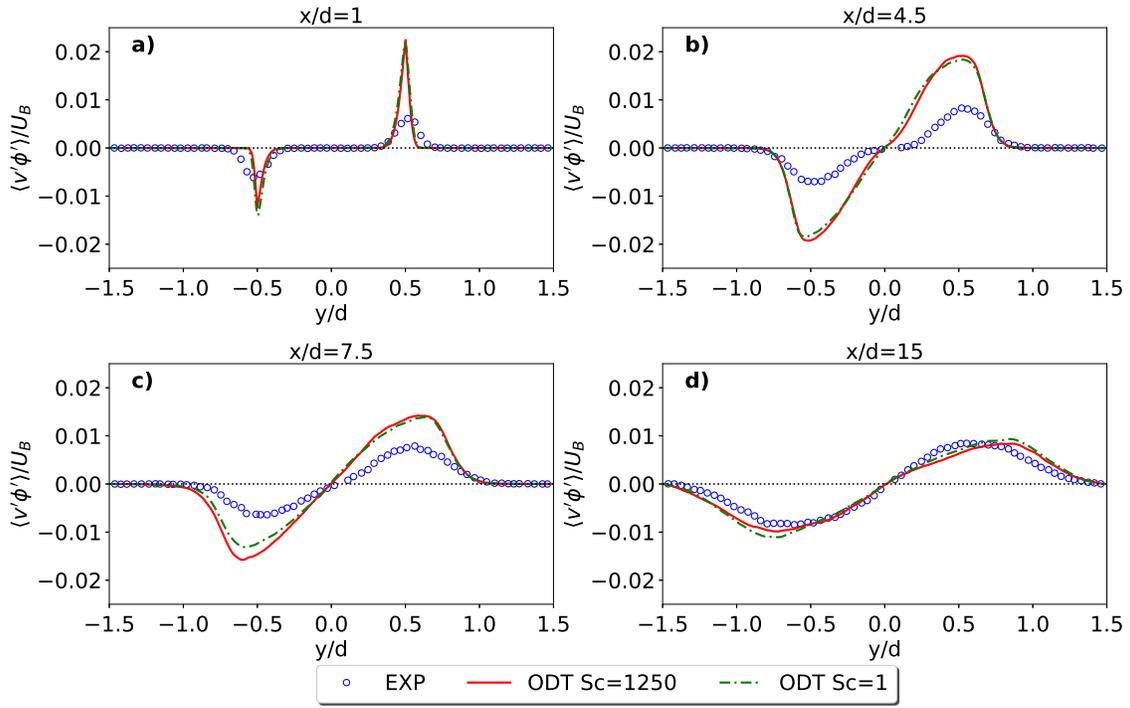}
  \caption{%
    Same as figure~\ref{img:Tmean_Feng} but for the wall-normal turbulent scalar flux $\langle v'\phi'\rangle$.
    The black dotted line gives the zero level.
    Measured reference data (EXP) are from~\cite{Feng2005}.
  }
  \label{img:vsFluc_Feng}
\end{figure}

\begin{figure}[p]
  \centering
  \includegraphics[scale=0.4]{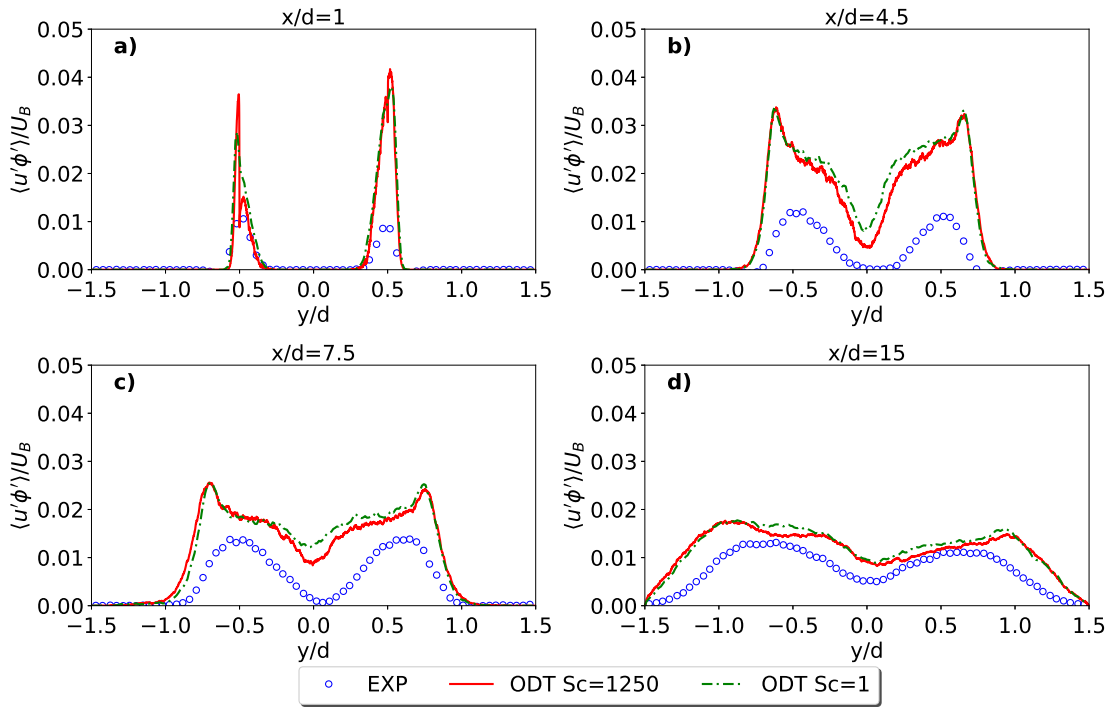}
  \caption{%
    Same as figure~\ref{img:vsFluc_Feng} but for the streamwise turbulent scalar flux $\langle u'\phi'\rangle$.
    Measured reference data (EXP) are from~\cite{Feng2005}.
  }
  \label{img:usFluc_Feng}
\end{figure}

Figures~\ref{img:vsFluc_Feng} and \ref{img:usFluc_Feng} show the wall-normal and streamwise turbulent scalar fluxes, respectively, for case~B with $Re=40\,000$, $Sc=1$ in comparison to $Sc=1250$.
The ODT results exhibit good qualitative agreement with the reference data of Feng et al.~\cite{Feng2005}.
Quantitative agreement is obtained only for the well-mixed downstream region at $x/d=15$, where the scalar is distributed more homogeneously across the channel so that its transport is reduced.

The difference between ODT and the reference data is largest near the inlet.
At $x/d=1$, both turbulent fluxes obtained with ODT are by a factor of $\approx3$--$4$ larger than the reference measurements.
This decreases to a factor of $\approx2$ at $x/d=4.5$ and $\approx1.5$ at $x/d=7.5$.
Further downstream, at $x/d=15$, the wall-normal scalar flux of the ODT results is in excellent agreement with the reference data but the streamwise one remains somewhat larger.
This suggests that small-scale resolution is critical for capturing the turbulent fluxes.

Note that the turbulent fluxes are virtually independent of the Schmidt number.
In fact, the streamwise turbulent scalar fluxes are equally bad for both $Sc=1$ and $1250$ precisely because of this independence.
This limitation of the model is, presumably, related to the modeling error of the velocity fluctuations $u'_{i,rms}$ (compare with figures~\ref{img:urms_Kong} and~\ref{img:vrms_Kong}).
Qualitatively, however, the Schmidt-number independence is consistent since the turbulent transport processes are expected to dominate over the molecular diffusive ones for $Sc\geq1$.

%%%%%%%%%%%%%%%%%%%%%%%
\subsection{Turbulence spectra of the streamwise velocity and the passive scalar} \label{sec:spectra}
%%%%%%%%%%%%%%%%%%%%%%%

With the ODT model it is feasible to compute high-resolution turbulence spectra for both the velocity and the passive scalar.
The model formulation yields continuous flow profiles for the ODT-resolved wall-normal direction at any instant in time or downstream location for that matter.
This is different for the streamwise direction, which exhibits discontinuities due to the sequence of mapping events (see figure~\ref{img:vis2d_Feng}).
The turbulence spectra of the streamwise velocity, $E_u(k_y)$, and the passive scalar, $E_\phi(k_y)$, are therefore computed for the wall-normal direction as described in \ref{sec:comp-spectra} so that $k_y$ denotes the wall-normal wave number.
Note that all ODT data has been interpolated to the same auxiliary grid to carry out the computation of the spectra and this grid resolves the Batchelor scale $\eta_B$ for $Sc=1250$.

The ODT-line interval and downstream location for the computation of the turbulence spectra are selected based on two criteria.
One is that the jet has had sufficient time to develop from its (synthetic) initial condition and another that the passive scalar is sufficiently distributed on the integral scale.
These requirements are fulfilled for the wall-normal interval $-0.5\leq y/d\leq 0.5$ at the downstream location $x/d=7.5$ (see figures~\ref{img:scheme_reactor} and \ref{img:vis2d_Feng}) for case~B with $Re=40\,000$ and $Sc\in\{1,\,1250\}$ (see table~\ref{tab:Allocation_reference}).
These spectra exhibit various similarity scalings in the form of power laws,
\begin{equation}
 E(k_y) \propto k_y^{-\gamma},
 \label{eq:E-power}
\end{equation}
where $\gamma$ is the scaling exponent; the subscripts $u$ and $\phi$ have been suppressed for readability.
These similarity scalings are discussed next.

\begin{figure}[t]
  \centering
%  %%%% >>>
%  \ \\[4ex] !! ATTENTION: FIG 15 ATTACHED AS TIFF !! \\[4ex] 
%  %%%% <<<
  \includegraphics[scale=0.38]{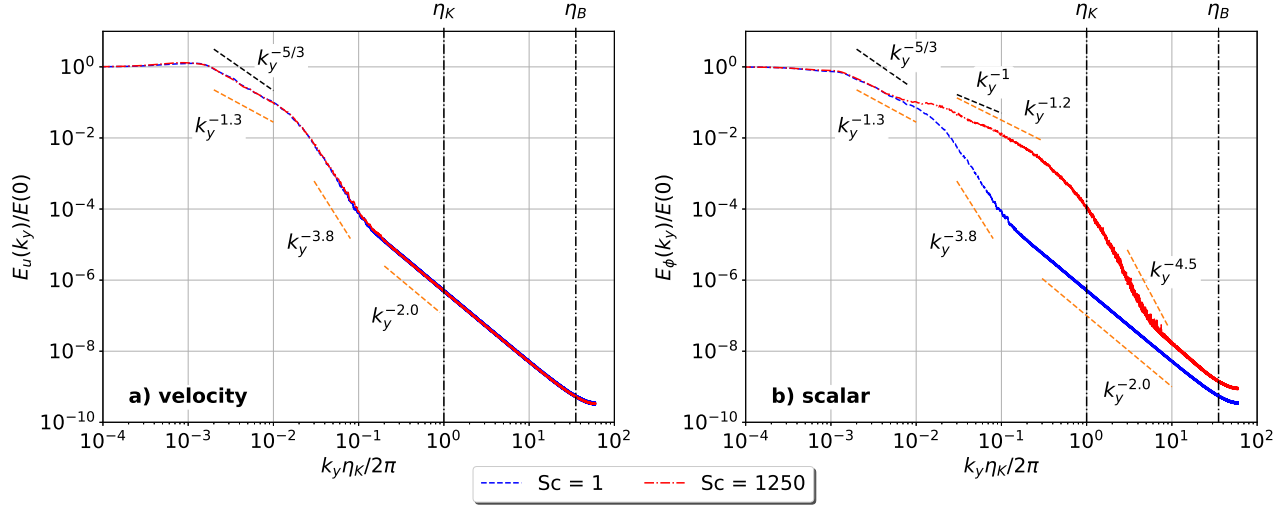}
  \caption{%
    One-dimensional turbulence spectra of the streamwise velocity $E_u(k_y)$ (a) and the passive scalar $E_\phi(k_y)$ (b) for case~B with $Re=40\,000$, $Sc=1$ and $1250$.
    All spectra have been obtained for the downstream location $x/d=7.5$ for the wall-normal interval $-0.5\leq y/d\leq0.5$ (see figure~\ref{img:scheme_reactor}). 
    Black vertical lines give the wave numbers $k_y$ that are equivalent to the Kolmogorov~($\eta_K$) and Batchelor~($\eta_B$) scales.
    Theoretical scaling laws according to Kolmogorov~\cite{Kolmogorov:1941} and Batchelor~\cite{Batchelor:1959} are given by black dashed lines.
    Empirical scaling laws of the ODT solution are given by orange dashed lines.
  }
  \label{img:spectra}
\end{figure}

Figure~\ref{img:spectra}(a) shows the turbulence spectra of the streamwise velocity.
As expected, the spectra are identical for both Schmidt numbers indicating that the dynamical grid adaption is not causing spurious results.
For small wave numbers (large scales), a Kolmogorov-like scaling can be discerned but the scaling exponent $\gamma_u=1.3$ is somewhat smaller than $5/3\approx1.67$, which would be expected for homogeneous isotropic turbulence~\cite{Kolmogorov:1941}.
This scaling does not continue down to the Kolmogorov wave number $2\pi/\eta_K$ in the present ODT results for the confined jet.
Instead, it is replaced by a much steeper fall-off with the exponent $\gamma_u=3.8$.

A very similar behavior has been observed by Giddey et al.~\cite{Giddey_etal:2018} in the case of forced isotropic turbulence.
These authors have shown that such a spectral drop in the inertial range is related to the suppression of small eddy events by a large value of the model parameter $Z$.
Having selected $Z=400$ here because of the boundary layer, the smallest eddy events are approximately $\sqrt{Z}\eta_K=20\eta_K$ in size.
This is a weak cut-off criterion but the corresponding cut-off wave length falls right in the $3.8$-scaling range. 
The conventional velocity statistics shown above are not notably modified by this cut-off since the affected scales carry at least two orders of magnitude less energy.
For very larger wave numbers (very small scales), the exponent $\gamma_u=2.0$ stems from the lower-order (1-D) diffusion and the discretization errors including interpolations used for the computation of the spectra.

Figure~\ref{img:spectra}(b) shows one-dimensional turbulence spectra of the passive scalar for $Sc=1$ and $1250$, respectively.
At low wave numbers, both spectra are very similar and exhibit the scaling exponent $\gamma_\phi=1.3$, which is the same as for the velocity and slightly smaller than the exponent $5/3\approx1.67$ of homogeneous, isotropic turbulence~\cite{Kolmogorov:1941} and is in very good agreement with the literature.
For passive scalars in sheared turbulence, the exponent $\gamma_\phi$ approaches the value $5/3$ from below with increasing Reynolds number~\cite{Warhaft:2000}.
Next, for somewhat larger wave numbers but still in the inertial-convective range, the scalar turbulence spectrum for $Sc=1250$ diverts from that for $Sc=1$. 
The scaling exponent reduces approximately to $\gamma_\phi=1.2$, which is slightly larger than unity of the Batchelor spectrum~\cite{Batchelor:1959}.
Nevertheless, the present ODT results for the high-$Re$ high-$Sc$ confined jet agree fairly well with the literature on homogeneous isotropic and sheared turbulence (e.g.~\cite{Donzis_etal:2010, Gotoh_Watanabe:2012}).
These authors, among others, have reported the coexistence of Kolmogorov and Batchelor spectra, but for different wave numbers, provided that the Reynolds number is large enough.
Here, a bulk Reynolds number of $40\,000$ is large enough in that sense.

Note that in the viscous-convective range, between the Kolmogorov and Batchelor wave numbers, the ODT turbulence spectrum for the $Sc=1250$ scalar turns gradually towards a $4.5$-scaling.
This is even steeper than the $3.8$-scaling of the $Sc=1$ scalar and likely related to the lower-order formulation.
Finally, in the viscous-diffusive range, the $2.0$-scaling is established similar to the velocity.

To summarize, the turbulence spectra obtained with ODT suggest that the assumption of locally isotropic turbulence is not fully valid for the confined jet.
Nevertheless, it is a very reasonable approximation which explains the success of gradient-diffusion closures for this case~\cite{Kong2012, Feng2005}.
The two-point velocity correlations shown by Kong et al.~\cite{Kong2012} support this interpretation as they are localized but only approximately symmetric in the wall-normal and streamwise directions.

%%%%%%%%%%%%%%%%%%%%%%%
\section{Conclusion} \label{sec:conc}
%%%%%%%%%%%%%%%%%%%%%%%

The turbulent mixing in confined jets is a canonical problem representative of a variety of chemical engineering applications.
We applied the lower-order, one-dimensional turbulence (ODT) model as stand-alone tool to resolve the flow down to the Kolmogorov and Batchelor scales.
The ODT computational domain is a wall-normal line, which is advected downstream with the bulk velocity during a simulation run.
In order to make ODT results comparable to the reference data, a multi-scale filtering approach has been adopted to synthesize inflow conditions.
These conditions were statistically consistent with the reference data of Kong et al.~\cite{Kong2012} and Feng et al.~\cite{Feng2005}.

In general, the results obtained indicate that ODT is able to capture important features of the confined jet.
This concerns, most importantly, the mean quantities and the turbulent fluxes along the computational domain.
For every downstream location in the case of a passive scalar with Schmidt number close to unity or when only the velocity is considered, ODT results agree, within a reasonable extent, with the coarsely resolved RANS~\cite{Feng2005}, LES~\cite{Kong2012} and LES-LEM~\cite{Arshad_etal:2018} reference data. 
Some (3-D) flow structures, however, remain unresolved by the lower-order model formulation.
This is a known limitation and manifests itself by up to $30\%$ underestimation of the r.m.s.~velocity fluctuations (e.g.~\cite{Kerstein:1999, Lignell_etal:2013}).
Even though expected otherwise, this modeling error is not visible for the scalar variance and present ODT results exhibit good agreement with the reference data for $Sc=1$.
For a high Schmidt number ($Sc=1250$), small-scale resolving ODT simulations exhibit much higher values of scalar variance, especially near the inlet (see figure~\ref{img:Trms_Feng}).
These values may even reach a factor $10$ disagreement with the coarsely resolved reference data.
This hints at implicit filtering effects in the reference data for the higher Schmidt number case.
For a Schmidt number close to unity ($Sc=1$), little can be gained by more expensive, or more finely resolved, approaches.

The interpretation, that filtering effects are present in the reference data, is supported by the ODT scalar turbulence spectra (figure~\ref{img:spectra}).
Only the ODT results for $Sc=1250$ exhibit a Batchelor-like spectrum for the wave numbers $k_y\eta_K/(2\pi)\geq10^{-2}$ (small scales).
This is in agreement with the literature on sheared and isotropic scalar turbulence~\cite{Warhaft:2000}.
The Bachelor spectrum is not even a decade (factor $10$) separated from the integral scale with respect to both wave number and magnitude.
This suggests that the filtering of the Batchelor spectrum by a coarse grid will easily modify the scalar fluctuations by $10$--$20\%$ at the downstream location $x/d=7.5$.
Closer to the inlet, this effect seems to be even larger. 
For the wave numbers $k_y\eta_K/(2\pi)\leq10^{-2}$ (large scales), the scalar turbulence spectra are independent of the Schmidt number and exhibit a Kolmogorov-like spectrum which is in very good agreement with the literature~\cite{Donzis_etal:2010, Gotoh_Watanabe:2012}.

Altogether, filtering effects, the increase of scalar fluctuations and the turbulence spectra, should be addressed in the future using both high-resolution measurements and high-resolution 3-D numerical simulations.
One potential candidate for this challenging computational task might be ODTLES, in which several ODT domains are physically coupled by the large-scale dynamics~\cite{Gonzalez-Juez_etal:2011, Glawe_etal:2018}.
With this 3-D expansion of the model, a better large-scale structure representation, which is missing in the stand-alone application of ODT, can therefore be achieved.

%%%%%%%%%%%%%%%%%%%%%%%
\section*{Acknowledgements} \label{sec:ack}
%%%%%%%%%%%%%%%%%%%%%%%

We thank two anonymous reviewers for their constructive criticism, which has helped to improve the quality of this manuscript.
We also thank Juan Ali Medina Mendez for commenting on the manuscript and discussions.

Funding: This work has been partly supported by the European Regional Development Funds (EFRE), Grant No.\ StaF~23035000; the German Academic Exchange Service (DAAD), which is funded by the Federal Ministry of Education and Research (BMBF), Grant No.\ PPP-USA-2017 (ID-57316240); and by a Mobility Grant ``Research Stay Abroad'' from the Graduate Research School (GRS) of the BTU Cottbus-Senftenberg. %%, Cluster~3, Stochastic Methods for Fluid Flow and Transport Processes.

%%%%%%%%%%%%%%%%%%%%%%%
\appendix
%%%%%%%%%%%%%%%%%%%%%%%

%%%%%%%%%%%%%%%%%%%%%%%
\section{Ensemble and temporal statistics} \label{sec:stats}
%%%%%%%%%%%%%%%%%%%%%%%

The variability of a transient flow, like the confined jet, is captured by an ensemble of $N$ ODT realizations.
These are realized as ensemble of initial conditions (see section~\ref{sec:ic} for details).
Pointwise ensemble statistics are computed for predefined times~$t^m$ or, correspondingly, downstream locations~$x(t^m)$.
Each ODT realization has its own grid so that the instantaneous profiles are interpolated to a common, equidistant, grid using a cubic spline interpolation~\cite{Press_etal:2007}.
This grid resolves the Kolmogorov scale or, if required, even the Batchelor scale.
The ensemble average $\langle \psi \rangle$ and variance $\langle \psi^{\prime\,2} \rangle$ are defined as
\begin{align}
  \langle \psi(y,t) \rangle &= \frac{1}{N} \sum_{n=1}^{N} \psi_n(y,t),
  \label{eq:mean}
  \\
  \langle \psi^{\prime\,2}(y,t) \rangle &= \frac{1}{N} \sum_{n=1}^{N} \big[ \psi_n(y,t) - \langle \psi(y,t) \rangle \big]^2 .
  \label{eq:var}
\end{align}
For the velocity vector, $\psi=u_i$, it is common to consider the standard deviation (root-mean-square fluctuations) $u'_{i,rms}=\sqrt{\langle u_i^{\prime 2}\rangle}$ rather than the variance.

Next, the Reynolds stress tensor $\langle u_i'u_j'\rangle$ and the turbulent scalar flux $\langle\phi'u_i'\rangle$ are resolved along the ODT line.
There, the triplet map (equation~\eqref{eq:triplet}) takes the role of an advecting velocity but for the duration of the corresponding eddy time-scale (equation~\eqref{eq:eddyTau}).
The turbulent fluxes are, therefore, computed by splitting the map-based from the diffusive transport.
This operation can \emph{not} be done alone with instantaneous property profiles but requires a small averaging time interval~$\Delta t_e$ over which eddy statistics are gathered separately.

The stochastic eddy event terms in equations~\eqref{eq:gov1} and \eqref{eq:gov2} represent a flux divergence.
In the lower-order formulation, this divergence is given by 
\begin{equation}
  \frac{\partial}{\partial y} \big(v'\psi'\big)(y,t) \approx -\frac{\Delta \psi_e}{\Delta t_e}(y,t),
 \label{eq:reystress-div}
\end{equation}
where $v'$ would be the fluctuating, physical, velocity component in the direction of the ODT line.
Furthermore,
\begin{equation}
 \Delta \psi_e(y,t) = \sum_{k=1}^{N_e} \big[ \psi''(y,t) - \psi(y,t) \big]_k
 \label{eq:reystress-du}
\end{equation}
is the accumulation of the map-induced (``turbulent'') changes of $\psi$ due to $N_e$ eddy events over the considered time interval $\Delta t_e$.
Note that $\psi''$ represents the map-induced changes $\phi''$ or $u_i''$ as defined by equations~\eqref{eq:eddy1} and \eqref{eq:eddy2}.
The ensemble-averaged flux divergence is then integrated in space to give the turbulent flux of $\psi$ along the ODT line coordinate as
\begin{equation}
 \langle v'\psi'(y,t) \rangle \simeq
   - \int^y
     \left\langle \frac{\Delta \psi_e}{\Delta t_e}(y',t) \right\rangle \,\mathrm{d}y'
 \label{eq:reystress}
\end{equation}

For the confined jet (see figure~\ref{img:scheme_reactor}), the averaging time interval $\Delta t_e$ seen in equation~\eqref{eq:reystress} is centered at the output time $t^m$ (which corresponds to the output location $x(t^m)$; see equation~\eqref{eq:t2x} for details).
This was done to reduce the bias from the transient flow in the spreading jet.
The size of $\Delta t_e$ is, thus, a compromise between a well-defined downstream location, computational efficiency (number of eddy events), and a reasonable estimation of the turbulent fluxes per realization.
Here, $\Delta t_e \, U_B/d \simeq 0.2$ has been selected corresponding to $\approx2\%$ of the total simulation time.

%%%%%%%%%%%%%%%%%%%%%%%
\section{Computation of one-dimensional turbulence spectra} \label{sec:comp-spectra}
%%%%%%%%%%%%%%%%%%%%%%%

One-dimensional turbulence spectra are obtained in a conventional way (see e.g.~\cite{Pope:2000}).
The direction is predefined by the orientation of the ODT-line coordinate $y$~\cite{Kerstein:1999, Giddey_etal:2018}.
For an ensemble of $N$ ODT realizations, the procedure is as follows:
\begin{enumerate}
 \item For a given variable $\psi=\phi,u$, extract the data at downstream location $x/d=\text{const}$ for a given ODT-line interval $[y_0,y_1]$ and interpolate to a common equidistant grid.
 \item Compute the fluctuations $\psi'$ for each member using ensemble statistics (\ref{sec:stats}).
 \item Compute the two-point spatial correlation $r_\psi(\tilde{y})$ of the fluctuations for each member,
 \begin{equation}
  r_{\psi}(\tilde{y}) = \int_{y_0}^{y_1} \psi'(y) \, \psi'(y+\tilde{y})\,\mathrm{d}y .
  \label{eq:corr1}
 \end{equation}
 \item Compute the autocorrelation $R_\psi(\tilde{y})$ for the whole ensemble,
 \begin{equation}
  R_\psi(\tilde{y}) 
    = \big\langle r_{\psi}(\tilde{y}) \big\rangle \Big/ \big\langle r_{\psi}(0) \big\rangle .
  \label{eq:corrN}
 \end{equation}
 \item Fourier-transform the (symmetric) autocorrelation which yields the one-dimensional turbulence spectrum,
 \begin{equation}
  E_\psi(k_y) = \dfrac{2}{\pi} \int_{-\infty}^{\infty} R_\psi(\tilde{y})\,\cos(k_y\,\tilde{y})\,\mathrm{d}\tilde{y} .
  \label{eq:spec}
 \end{equation}
\end{enumerate}

%%%%%%%%%%%%%%%%%%%%%%%
\section{Turbulent inflow conditions obtained with ODT} \label{sec:ic-odt}
%%%%%%%%%%%%%%%%%%%%%%%

In the reference experiments~\cite{Feng2005, Kong2012, Arshad_etal:2018} the confined planar jet is realized by bringing together three turbulent duct flows with a large aspect ratio (see figure~\ref{img:scheme_reactor}).
The corresponding LES use synthetic inflow conditions that are statistically consistent with the measurements as described in section~\ref{sec:ic}.
In the case of ODT, the ``natural'' approach is a self-consistent one.
That is, the turbulent inflow conditions are generated with ODT using three incompressible fully-developed turbulent channel flow solutions for which the model set-up is well-documented (e.g.~\cite{Lignell_etal:2013}).
Here, however, it turned out that this approach is unable to provide inflow conditions that are consistent with the reference data.  
In the following, we discuss the ``natural'' approach nevertheless as it can be useful for other engineering applications that demand the prescription of a reasonable turbulence field.

In order to generate turbulent inflow conditions for the confined jet with ODT, each small channel of size~$d$ has been simulated separately with the model parameters $C=10$, $Z=600$, $\alpha=2/3$~\cite{Lignell_etal:2013}.
The symmetry properties of the fully-developed flow are addressed by a fraction-of-domain large-eddy suppression method, which has been used instead of the elapsed-time method that was used for the confined jet (see section~\ref{sec:model} and table~\ref{tab:Allocation_reference}). 
That is, the suppression parameter is set to zero ($\beta_{LS}=0$) and the maximum allowed size of the eddy events is $l_{max}=d/2$.

The individual ODT channel flow simulations are conducted analogous to the case set-up described in section~\ref{sec:config} but the transient stage is neglected.
Only one long-time simulation is needed for the central jet and the co-flow to form two ensembles of inflow profiles taken from the statistically stationary state.
(Statistical stationarity has been confirmed by monitoring the temporal averages described in \ref{sec:stats} for several thousand eddy events.)
Flow profiles from the small channels are randomly selected from these ensembles and combined to form an ensemble of turbulent inflow conditions, which is then prescribed for the large channel at $x/d=0$.
A small region $\Delta y/d\ll1$ was cut around each splitter plate ($y/d=\pm0.5$) to account for their finite thickness as visible in the reference data~\cite{Feng2005, Kong2012}.

\begin{figure}[t]
  \centering
  \includegraphics[scale=0.5]{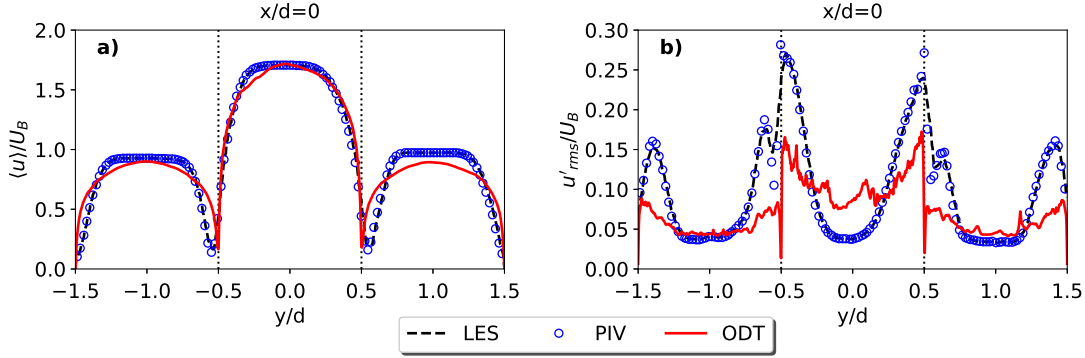}
  \caption{
    Initial streamwise mean velocity~(a) and r.m.s.~velocity fluctuations~(b) for case~A by combination of three ODT channel flow simulations using $N=5000$ flow profiles.
    Statistical moments of the reference PIV and initial LES data are on top of each other.
    Dotted vertical lines give the locations of the splitter plates for orientation.
  }
  \label{img:ODT_initial_condition_Kong}
\end{figure}

Figure~\ref{img:ODT_initial_condition_Kong} shows the mean velocity $\langle u \rangle/U_B$~(a) and root-mean-square (r.m.s.) fluctuations $u'_{rms}/U_B$~(b) of the initial conditions generated with ODT.
The corresponding reference data~\cite{Kong2012} (symbols) and locations of the splitter plates (broken vertical lines) are also given.
The ODT mean velocity profile exhibits reasonable agreement with the reference data, but the flat section in each center of the three channel flows, the almost linear sections towards the walls, and the slight left-right asymmetry of the reference data can \emph{not} be reproduced.
Especially the ODT r.m.s.~velocity fluctuations of the turbulent inflow are \emph{not} adequately capturing the reference data.
The modeling error is a factor of two, which is strongest around the splitter plates ($y/d=\pm0.5$).
These differences result from secondary flows that are generated in the inlet section~\cite{Feng2005, Kong2012} and the reference duct geometry~\cite{Vinuesa_etal:2018, Modesti_etal:2018}.
Therefore and in order to avoid initially biased results, we have implemented a synthetic turbulent inflow condition as described in section~\ref{sec:ic}.

%%%%%%%%%%%%%%%%%%%%%%%
\section*{References}
%%%%%%%%%%%%%%%%%%%%%%%

% %%% >>>
% \bibliography{mybibfile}
% %%% <<<

%%%%%%%%%%%%%%%%%%%%%%%
\end{document}